\documentclass[aps,prl,twocolumn,superscriptaddress,preprintnumbers,floatfix,10pt,nofootinbib]{revtex4-2}

\usepackage[bookmarks=false]{hyperref}
\usepackage{amsmath}
\usepackage{graphicx}
\usepackage{xspace}
\usepackage[usenames,dvipsnames]{xcolor}

\DeclareRobustCommand{\refcite}[1]{Ref.~\cite{#1}}

\DeclareRobustCommand{\eq}[1]{Eq.~\eqref{eq:#1}}

\DeclareRobustCommand{\fig}[1]{\hyperref[fig:#1]{Fig.~\ref*{fig:#1}}}

\newcommand{\nn}{\nonumber}

\newcommand{\Mae}[3]{\big\langle#1\big\lvert#2\big\rvert#3\big\rangle}

\newcommand{\bn}{{\bar n}}

\newcommand{\cE}{\mathcal{E}}

\newcommand{\cJ}{\mathcal{J}}

\newcommand{\as}{\alpha_s}

\newcommand{\lqcd}{\Lambda_\mathrm{QCD}}
\newcommand{\MSbar}{$\overline{\text{MS}}$}

\newcommand{\tr}{\operatorname{tr}}

\newcommand{\Pythia}{\texttt{Pythia}\xspace}
\newcommand{\Herwig}{\texttt{Herwig}\xspace}

\newcommand{\N}{[N]}
\newcommand{\xL}{x_{L}}

\allowdisplaybreaks[2]

\providecommand{\sectionPaper}[1]{\textit{\textbf{#1.}}}
\providecommand{\headingAcknowledgments}{\textit{\textbf{Acknowledgments. }}}
\providecommand{\citeSupplementBibEntry}{\cite{*[{See Supplemental Material at \textcolor{red}{\texttt{<insert url>}}}] [{}] supplement}}
\providecommand{\app}[1]{\citeSupplementBibEntry}
\providecommand{\apps}[2]{\citeSupplementBibEntry}

\begin{document}

\preprint{\vbox{\hbox{MIT--CTP 5711}}}
\preprint{\vbox{\hbox{DESY-24-064}}}

\title{Nonperturbative Effects in Energy Correlators: \\
From Characterizing Confinement Transition to Improving $\boldsymbol{\alpha_s}$ Extraction}

\author{Kyle Lee}%
\email{kylel@mit.edu}%
\affiliation{Center for Theoretical Physics, Massachusetts Institute of Technology, Cambridge, MA 02139, USA}%

\author{Aditya Pathak}%
\email{aditya.pathak@desy.de}%
\affiliation{Deutsches Elektronen-Synchrotron DESY, Notkestr. 85, 22607 Hamburg, Germany}%

\author{Iain W.~Stewart}%
\email{iains@mit.edu}%
\affiliation{Center for Theoretical Physics, Massachusetts Institute of Technology, Cambridge, MA 02139, USA}%

\author{Zhiquan Sun}%
\email{zqsun@mit.edu}%
\affiliation{Center for Theoretical Physics, Massachusetts Institute of Technology, Cambridge, MA 02139, USA}%

\date{September 30, 2024}

\begin{abstract}
Energy correlators provide a powerful observable to study fragmentation dynamics in QCD.
We demonstrate that the leading nonperturbative corrections for projected $N$-point energy correlators are described by the same universal parameter for any $N$, which has already been determined from other event shape fits.
Including renormalon-free nonperturbative corrections substantially improves theoretical predictions of energy correlators, notably the transition into the confining region at small angles.
Nonperturbative corrections are shown to have a significant impact on $\alpha_s$ extractions.

\end{abstract}

\maketitle

\sectionPaper{Introduction}
The mysterious process of fragmentation in Quantum Chromodynamics  (QCD)
describes how an individual quark or gluon undergoes hadronization to turn into hadronic final states in particle colliders.
The angular distribution of energy in high-energy scattering offers a powerful experimental lens through which to explore the Lorentzian dynamics of this fragmentation process. Central to this exploration is the energy flow operator~\cite{Sveshnikov:1995vi,Tkachov:1995kk,Korchemsky:1999kt,Bauer:2008dt,Hofman:2008ar,Belitsky:2013xxa,Belitsky:2013bja,Kravchuk:2018htv},
\begin{align}
\label{eq:EnergyFlow}
\mathcal{E}(\vec{n})=\int_0^{\infty} d t \lim _{r \rightarrow \infty} r^2 n^i T_{0 i}(t, r \vec{n})\,,
\end{align}
which relates the energy-momentum tensor $T_{\mu\nu}$ in QCD to the energy captured by a detector in the direction $\vec n$. Measuring correlators of these operators
$\left\langle\Psi\left|\mathcal{E}\left(\vec{n}_1\right) \mathcal{E}\left(\vec{n}_2\right) \cdots \mathcal{E}\left(\vec{n}_N\right)\right| \Psi\right\rangle$
through $N$-point energy correlators (ENC),
enables us to uncover quantum correlations in collider energy distributions~\cite{Basham:1979gh,Basham:1978zq,Basham:1978bw,Basham:1977iq,Dixon:2019uzg,Chen:2020vvp}.
The state $|\Psi\rangle$ could be generated by a local source such as an electromagnetic current injecting invariant mass $Q^2$,
or denote a collection of particles such as a jet~\cite{Lee:2022ige}.
Varying the angles $z_{ij} = \big(1- \vec n_i \cdot \vec n_j\big)/2$
reveals numerous intrinsic and emergent scales in QCD~\cite{Barata:2023bhh,Komiske:2022enw,Holguin:2022epo,Lee:2022ige,Liu:2022wop,Liu:2023aqb,Cao:2023oef,Devereaux:2023vjz,Andres:2022ovj,Andres:2023xwr,Craft:2022kdo,Lee:2023npz,Holguin:2023bjf}.
By exploring the energy correlations across the entire scattering event --- similar to cosmic surveys mapping the sky --- we can neatly map particle interactions at their most fundamental level. This also provides a unique probe of a key parameter of the Standard Model, the strong coupling $\as$.

\begin{figure}[t!]
\begin{center}
\includegraphics[scale=0.38]{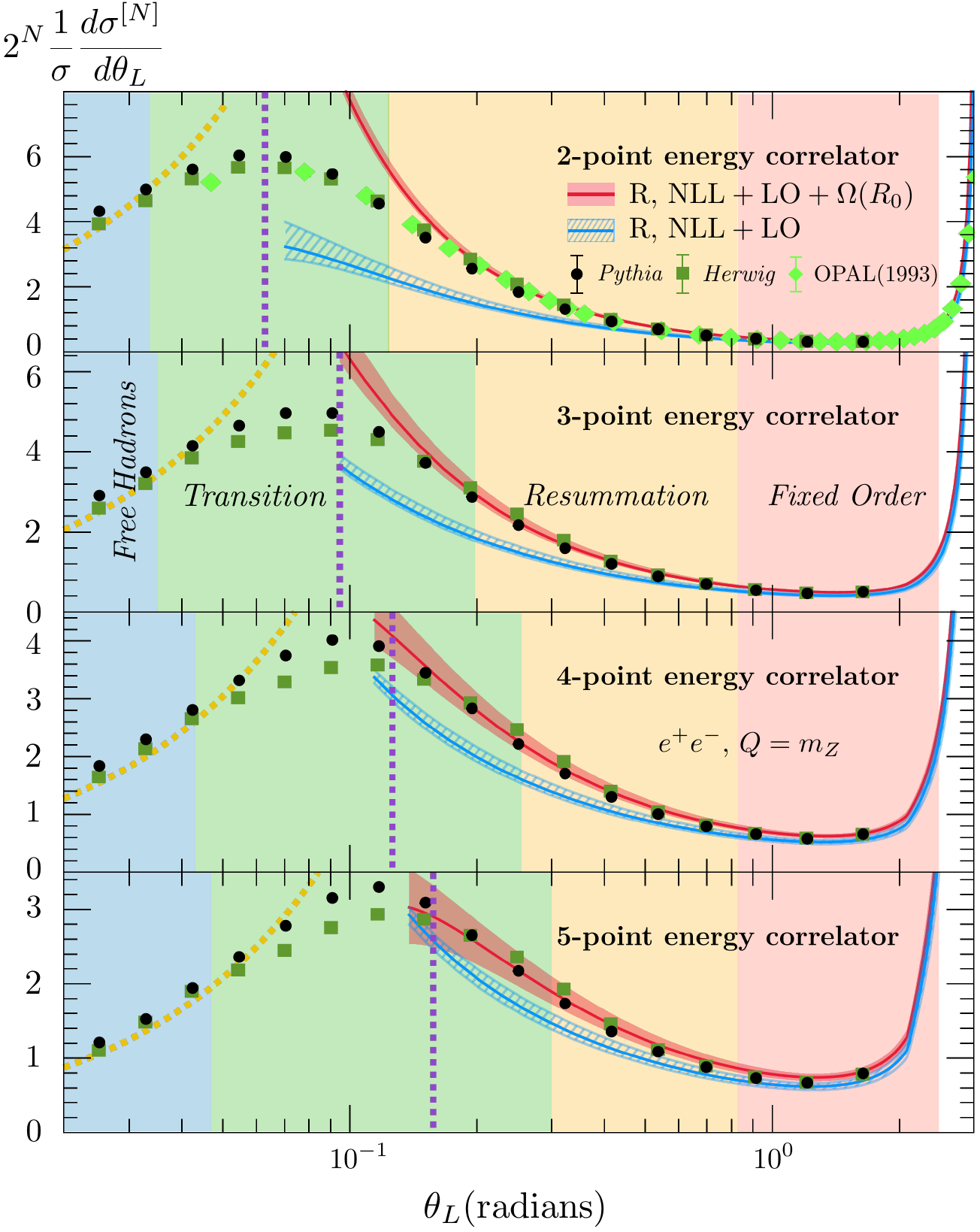}
\end{center}
\vspace{-0.4cm}
\caption{The projected $N$-point energy correlators for $N=2$--$5$ at $Q=m_Z$ display a transition from the perturbative to free hadron scaling region for small angles.
The R scheme calculation with $\alpha_s$ and $\Omega_1$ values from~\refcite{Abbate:2010xh} significantly improves the description near this transition without fitting.
\vspace{-0.45cm}}
\label{fig:transition}
\end{figure}

In this \emph{Letter}, we determine the leading nonperturbative effects for projected ENCs (pENCs)\cite{Chen:2020vvp}, which measure only the largest angle $x_L \equiv (1-\cos\theta_L)/2=\mathrm{max}\{z_{ij}\}$.
We find they are determined by two universal hadronic parameters, $\Omega_{1q}$ and $\Omega_{1g}$, with perturbatively calculable $N$ and $x_L$ dependence.
The same $\Omega_{1\kappa}$ appear in other observables, providing exciting prospects for testing universality and making parameter-free predictions.
Eliminating the  renormalon ambiguity in $\Omega_{1\kappa}$, we find an improved description of the transition from the partonic to confining region for pENCs, see \fig{transition}.
Motivated by the recent CMS measurement of $\alpha_s$, using a ratio of energy correlators~\cite{CMS:2024mlf}, we assess the impact of nonperturbative effects in ratios and the accuracy of estimates based on $e^+e^-$ Monte Carlo generators (MC).

The two-point energy correlator (EEC) is given by
\begin{align}
\frac{d\sigma^{[2]}}{dz} &= \sum_X \int d \sigma_{e^{+} e^{-} \rightarrow X} \sum_{i,j \in X} \frac{E_{i}E_{j}}{Q^2}\delta\left(z-\frac{1-\cos \theta_{ij}}{2}\right)
\nonumber\\
&= \int d^4 x \frac{e^{i q \cdot x}}{Q^2}
\int d \Omega_{\vec{n}_1} \int d \Omega_{\vec{n}_2}
\delta\left(z-\frac{1-\vec{n}_1 \cdot \vec{n}_2}{2}\right)
\nonumber\\
&L_{\mu\nu} \times\left\langle 0\left|J^{\mu\dagger}(x) \mathcal{E}\left(\vec{n}_1\right) \mathcal{E}\left(\vec{n}_2\right)J^\nu(0)\right| 0\right\rangle\,,
\end{align}
where $X$ refers to the hadronic final state sourced by the electromagnetic current $J^{\mu} = \overline \psi \gamma^\mu \psi$, $L_{\mu\nu}$ is the leptonic tensor, and $d\Omega_{\vec{n}_i} = \sin\theta_i d\theta_i d\phi_i / 4\pi$. The second equality gives us an operator definition of EEC in terms of a correlation function of energy flow operators.
Exploiting this operator definition has enhanced the interpretability of experimental data~\cite{CMS:2024mlf, Mazzilli:2024ots, Tamis:2023guc, Fan2023}, facilitating the application of a wide array of modern field-theoretic techniques, and inspired precision calculations~\cite{Chicherin:2020azt,Chen:2022jhb,Chang:2022ryc,Chen:2023wah,Alday:2016njk,DelDuca:2016csb, Tulipant:2017ybb,Dixon:2018qgp,Dixon:2019uzg, Korchemsky:2019nzm, Chen:2020uvt,Kodaira:1981nh, Kodaira:1982az, deFlorian:2004mp, Tulipant:2017ybb, Moult:2018jzp, Ebert:2020sfi,Duhr:2022yyp,Belitsky:2013xxa,Belitsky:2013bja,Belitsky:2013ofa, Henn:2019gkr, Moult:2019vou,Kologlu:2019mfz}.

\begin{figure}
\begin{center}
\includegraphics[scale=0.28]{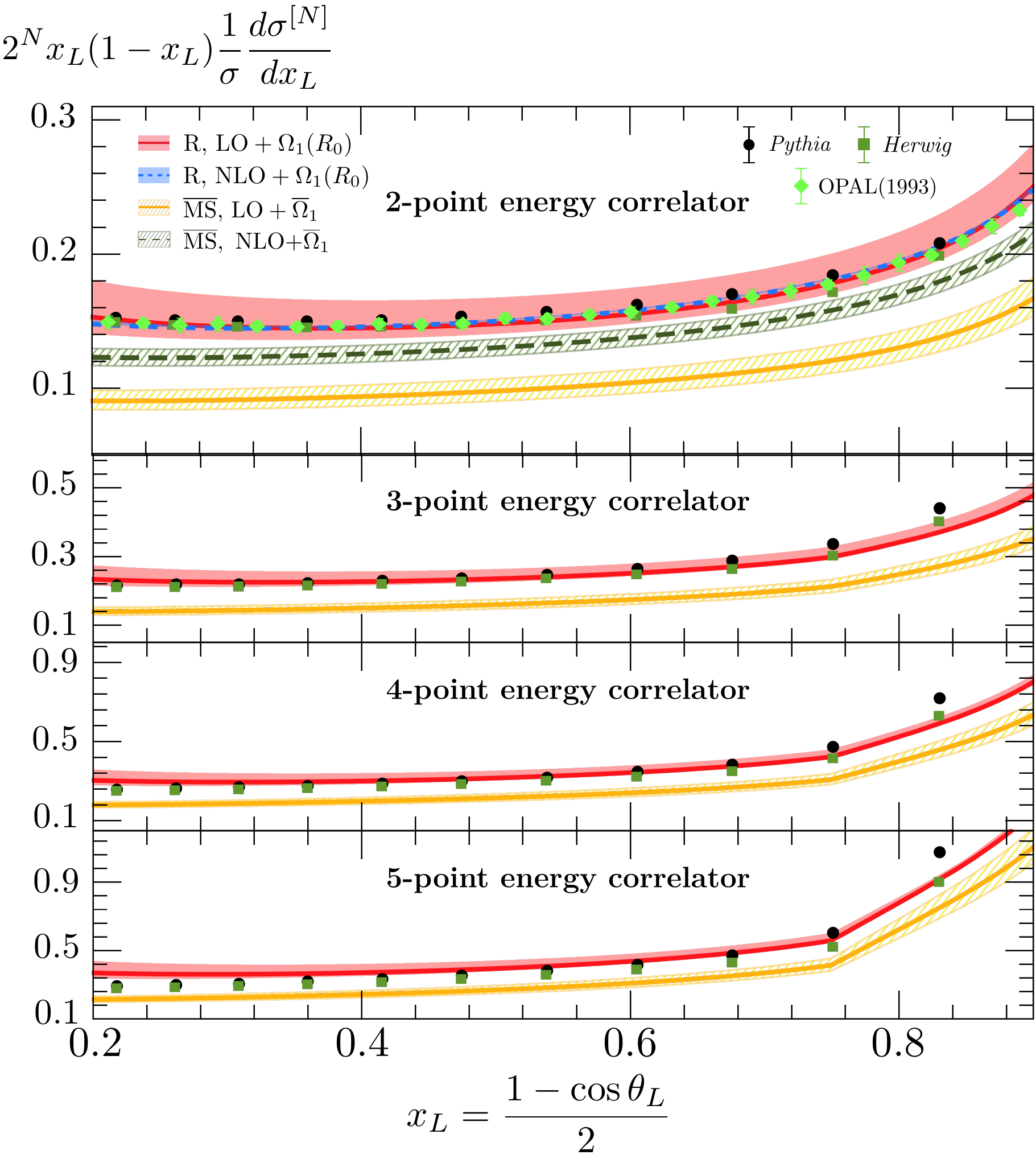}
\end{center}
\caption{The R scheme fixed-order calculation for pENCs provides a significant improvement compared to the $\overline{\mathrm{MS}}$ scheme, shown here away from the kinematic endpoints. For $N> 2$, a kink is located at $x_L =(1-\cos(2\pi/3))/2=3/4$, where the contributions probing three particles become relevant.
}
\label{fig:FO}
\end{figure}

Examining the effects of hadronization on such correlations has been one of the prime objectives of collider-QCD studies. In a pioneering paper~\cite{Korchemsky:1999kt}, the leading hadronization power corrections for EEC were first identified,  described by the matrix element~\cite{Lee:2006nr}
\begin{align}
\label{eq:Omega}
\Omega_{1\kappa} \equiv
\frac{1}{N_\kappa} \Mae{0}{ \tr \overline{Y}_{\bn}^{\dagger\kappa} Y_{n}^{\dagger\kappa}
\cE_T(0) Y_{n}^\kappa \overline{Y}_{\bn}^\kappa}{0}
\,, \end{align}
where $\kappa=q,g$ denotes the color channel~\cite{Stewart:2014nna,Ferdinand:2023vaf}, and $N_q = N_c$ and $N_g = N_c^2 -1$. The Wilson lines, $Y_{n(\bar{n})}^\kappa$, are oriented along back-to-back light-like directions $n^\mu$, $\bn^\mu$ and source the soft hadrons, and $\cE_T(\eta)$ is the transverse energy flow operator. Here $\Omega_{1q}$ is the same leading power correction appearing in $e^+e^-$ event shapes in the dijet limit~\cite{Dokshitzer:1995zt,Gardi:2000yh,Lee:2006nr,Abbate:2010xh,Mateu:2012nk}.
In~\refcite{Schindler:2023cww}, the leading renormalon ambiguity was calculated for the EEC,
and results for the EEC were presented eliminating this leading ambiguity, and utilizing a value for the leading power correction obtained from an earlier thrust fit~\cite{Abbate:2010xh}.
Away from kinematic endpoints, this reorganization addressed the long-standing discrepancies between perturbative calculations and $e^+e^-$ collider data~\cite{OPAL:1990reb,ALEPH:1990vew,L3:1991qlf,OPAL:1993pnw} for EEC without the need to fit any parameters as shown in the top panel of \fig{FO}.

The pENCs that we study are defined by
\begin{align}
\label{eq:pENC}
&\hspace{-0.2cm}\frac{d \sigma^{[N]}}{d x_L}
= \int\!\! d^4 x \frac{e^{i q \cdot x}}{Q^N}\prod_{i=1}^N\int d \Omega_{\vec{n}_i} \delta\left(x_L-\frac{1\!-\!\mathrm{min}\left(\vec{n}_i \!\cdot\! \vec{n}_j\right)}{2}\right) \nonumber\\
&\hspace{0.15cm} \times L_{\mu\nu}\left\langle 0\left|J^{\mu\dagger}(x) \mathcal{E}\left(\vec{n}_1\right) \mathcal{E}\left(\vec{n}_2\right)\ldots \mathcal{E}\left(\vec{n}_N\right) J^\nu(0)\right| 0\right\rangle .
\end{align}
Nonperturbative effects grow in significance as we approach the kinematic endpoints, $x_L\to 0,1$, where the invariant mass scale becomes nonperturbative, $Q\sqrt{x_L(1-x_L)}\sim \Lambda_{\rm QCD}$.
Here energy correlators exhibit a striking transition from the perturbative partonic scaling region to freely propagating confined hadrons with vanishing correlations.
This transition has been observed for $N=2$ and $N=3$ in jets with hadron collider data~\cite{CMS:2024mlf,Tamis:2023guc,Fan2023}.
For $e^+e^-$ this transition is clearly visible in \fig{transition} for $N=2-5$ in \Pythia and \Herwig simulations.

\sectionPaper{Universal power corrections}
Away from $x_L \to 1$ for pENCs from $e^+e^-$, the leading $\Lambda_{\rm QCD}/Q$ corrections at lowest order in $\alpha_s$  arise when one of the energy flow detectors aligns with the direction of soft hadrons, while the remaining $N-1$ detectors align with one of the energetic dijets.  Thus, for this contribution $x_L$ specifies the angle between the soft hadron direction and the energetic dijet, and there are $N$ different ways to place a detector on soft hadrons. Using $\cosh\eta = 1/(2\sqrt{x_L(1-x_L)})$ and $\cE_T(\eta) = \cosh^{-3}\!\eta \int d\phi\,\mathcal{E}(\vec{n})$, we arrive at the form of leading nonperturbative power corrections for pENCs from $e^+e^-$ collisions:
\begin{align}
\label{eq:pENCpc}
\frac{1}{\sigma}\frac{d\sigma^{[N]}}{dx_L} = \frac{1}{\sigma}\frac{d\hat{\sigma}^{[N]}}{dx_L} + \frac{N}{2^{N}}\frac{\overline{\Omega}_{1q}}{Q(x_L(1-x_L))^{3/2}}\,,
\end{align}
where $\hat{\sigma}^{[N]}$ is the perturbative result and $\overline{\Omega}_{1q} \sim \Lambda_{\rm QCD}$ is the nonperturbative power correction in the $\overline{\mathrm{MS}}$ scheme. The overall $1/2^N$ in the second term, also present in the first term, arises from the energy weighting $Q^N$ needed to ensure the sum rule
$\int_0^1 dx_L \: d\sigma^{[N]}/dx_L = \sigma$.
For $N=2$ this result was derived in~\refcite{Korchemsky:1999kt} with nonperturbative matrix element $\Omega_{1q}$ as defined in \eq{Omega}, and we extend this to $N>2$. For typical event shapes, an analogous relation for the leading power correction is only valid in the dijet limit. This is because the measurement in such event shapes is not inclusive and directly constrains the momentum of the final state. In contrast, energy correlations for any angular separation are inclusive, such that this relation remains valid across all angles except when $x_L\to 1$.
(In the back-to-back limit, contributions from detectors on three particles also give a leading contribution.)
At higher orders in $\alpha_s$ the $N$ and $x_L$ dependence can be different.

Next, we bring in insights from renormalon analysis of bubble-chain diagrams, where the leading $n_f$ terms at each perturbative order give a mechanism for examining the nature of power corrections in asymptotic perturbative series in $\overline{\mathrm{MS}}$~\cite{Beneke:1998ui,Argyres:2012ka, Dunne:2013ada}.
Renormalon analysis has various benefits: It provides an independent check on the coefficient of the nonperturbative matrix element,
including the $\xL$ and $N$ dependence in \eq{pENCpc}.
Also removing renormalon ambiguities present in  $\overline{\rm MS}$ from both $d\hat \sigma^{[N]}$ and $\overline{\Omega}^{[N]}_{1q}$  in \eq{pENCpc} improves the convergence of the series already at low orders in perturbation theory.
Furthermore, with the renormalon ambiguity removed, the parameters $\Omega_{1 \kappa}$ can now fully capture the leading nonperturbative effects.
The bubble sum calculation for pENCs is a simple extension of~\refcite{Schindler:2023cww}, and we find a `$u = 1/2$' pole in the Borel space~\cite{Gross:1974jv, Lautrup:1977hs, tHooft:1977xjm} with the ambiguity from the contour around this pole given by
\begin{align} \label{eq:amb_ENC}
&\Delta_{1/2} \bigg(2^N \frac{1}{\sigma_{0}}  \frac{d \hat\sigma ^{\N}}{d \xL}  \bigg)
=  \frac{N}{2} \Delta_{1/2} \bigg(2^2 \frac{1}{\sigma_{0}}  \frac{d \hat\sigma ^{[2]}}{d \xL}  \bigg)
\nn \\
&=  -  \frac{N}{2} \frac{8 i  C_{F} e^{5/6}}{\beta_{0}} \frac{2^2}{[\xL(1-\xL)]^{3/2}} \frac{\lqcd}{Q}
\,.
\end{align}
This confirms
the $x_L$ and $N$ dependence in \eq{pENCpc}.

We restore the separation of scales in power corrections by using the R scheme with a subtraction scale $R$~\cite{Hoang:2007vb,Hoang:2008fs,Hoang:2009yr,Bachu:2020nqn} to remove the renormalon ambiguities from both $\overline{\Omega}_{1\kappa}$ and ${d\hat{\sigma}^{[N]}}/{dx_L}$ in the \MSbar~scheme. This can be done by defining $\Omega_{1 \kappa}(R)$ and ${d\hat{\sigma}^{[N]}_{\rm R}}/{dx_L}$ as
\begin{align}
\label{eq:FORscheme}
\Omega_{1 \kappa}(R) & \equiv \overline\Omega_{1 \kappa} \!-\!  R \sum_{n=1}^\infty d_{\kappa n}\Bigl( \frac{\mu}{R}\Bigr) \bigg[ \frac{\alpha_s(\mu)}{4\pi}\bigg]^n
\,, \\
\frac{1}{\sigma}\frac{d\hat\sigma_\mathrm{R}^{\N}(R)}{d\xL}
&\equiv \!  \sum_{n=1}^\infty \bigg\{ c_n\Big(\xL, \frac{\mu}{Q}\Big)
\nn \\
&+ \frac{N}{2^N}
\frac{R}{Q} \frac{d_{qn}\bigl(\mu/R)\bigr)}{[\xL (1-\xL)]^{3/2}} \bigg\}
\bigg[  \frac{\alpha_s(\mu)}{4\pi} \bigg]^n
\nn
\,, \end{align}
where the coefficients $c_{n}$ are those of the original \MSbar ~series and $d_{\kappa n}$ are functions of logarithms $\ln(\mu/R)$ with coefficients appropriately chosen to remove the renormalon~\cite{Hoang:2007vb,Abbate:2010xh,Bachu:2020nqn,Hoang:2014wka,Abbate:2010xh}.  To ensure that $\Omega_{1 \kappa}(R)$ is of the same parametric size as $\Omega_{1 \kappa}$,
we generally choose $R$ such that the change is $\sim\lqcd$, and resum the large logarithms between $R$ and $\mu$ using $R$-RGE~\cite{Hoang:2008fs,Abbate:2010xh,Bachu:2020nqn}. Using the $\alpha_s = 0.114$ and $\Omega_{1q}^{\mbox{\tiny \cite{Abbate:2010xh}}}(R\!=\!2\,{\rm GeV})=0.323\,{\rm GeV}$ values determined from the thrust fit in \refcite{Abbate:2010xh}, we convert to the R scheme \texttt{\#}2 of \refcite{Dehnadi:2023msm}, and add 23\% from hadron mass corrections following~\cite{Mateu:2012nk,Schindler:2023cww}, yielding $\Omega_{1q}(R\!=\!2\,\rm{GeV})=0.658\,{\rm GeV}$.
We investigate the (small) dependence on the R scheme choice in a future paper~\cite{LPSS-Long}.

In \fig{FO}, we present fixed-order calculations at $\mathcal{O}(\alpha_s)$ at the $Z$-pole, $Q=m_Z$, incorporating power corrections for both the $\overline{\mathrm{MS}}$ and R schemes. Uncertainty bands are from factor of two scale variations around $\mu=Q$. We compare our results with MC simulations from \Pythia and \Herwig, and with OPAL data~\cite{OPAL:1993pnw} for $N=2$. For $N=2$, we observe improved perturbative convergence in the R scheme relative to the $\overline{\mathrm{MS}}$ scheme and a remarkable agreement with the OPAL data (consistent with \refcite{Schindler:2023cww} in a different R scheme). For higher-point cases, with a LO analysis, we find that the R scheme leads to better agreement with MC, similar to $N=2$. Thus we anticipate similar improvements in perturbative convergence for $N>2$, indicating that the R scheme is effective in removing the dominant renormalon ambiguity. This provides strong motivation for comparing R scheme-improved perturbative predictions with real-world collider data, particularly with upcoming revised LEP data  analyses on the horizon~\cite{Chen:2023nsi}.

\sectionPaper{Imaging Hadronization Transition}
The leading scaling behavior of pENCs in the small angle limit is captured through an iterative application of the light-ray operator product expansion (OPE)~\cite{Kologlu:2019mfz,Chang:2020qpj,Kravchuk:2018htv,Hofman:2008ar,Belitsky:2013xxa,Belitsky:2013bja,Caron-Huot:2022eqs} and is given by the twist-$2$ spin-$(N+1)$ anomalous dimension.
A QCD factorization theorem for pENC in the collinear limit, which captures both this anomalous scaling and the RG-flow of the coupling, was derived both for $e^+e^-$ colliders~\cite{Dixon:2019uzg} and for jets in hadron colliders~\cite{Lee:2022ige}. In the small angle limit, the cumulant of the pENC $\Sigma^{[N]}$, factorizes as
\begin{align}
\label{eq:convol}
&\Sigma^{[N]}(x_L) =\frac{1}{\sigma} \int^{x_L}_0 dx'_L \frac{d\sigma^{[N]}}{dx'_L}  \\
&=\int_0^1 dx\, x^N\, \vec{J}^{[N]}\left(\ln \frac{x_L x^2 Q^2}{\mu^2}, \mu\right) \cdot \vec{H}\left(x, \frac{Q^2}{\mu^2}, \mu\right)\,,\nn
\end{align}
where $\vec{H}$ denotes a hard function that describes the production of the energetic particles, and $\vec{J}^{[N]}$ is the $N$-point energy correlator jet function sensitive to the $x_L$ dependence of the observable ($\vec{J}^{[N]}$ is known at next-to-next-to-leading logarithmic order for $N=2,3$~\cite{Dixon:2019uzg,Chen:2023zlx}). Both functions are vectors in the gluon/quark space.
For hadron colliders, $\vec H$ also includes parton distribution functions.
The evolution equation for the jet function is
\begin{align}
\label{eq:Jevol}
\frac{d \vec{J}^{[n]}\Bigl(\ln \frac{x_L Q^2}{\mu^2}\Bigr)}{d \ln \mu^2}
=\! \int_0^1\!\!\! d y\,y^n\,\vec{J}^{[n]}\Bigl(\ln \frac{x_L y^2 Q^2}{\mu^2}\Bigr) \cdot \widehat{P}(y)\,,
\end{align}
where $\widehat{P}(y)$ is the singlet time-like splitting matrix.

As the renormalon ambiguity plagues pENCs at all angles, including the small-angle region, it must also be removed from these jet functions. Demanding consistency with the RG and the prediction in \eq{pENCpc} for nonperturbative corrections in the fixed order region, the power corrections to the jet functions must take the form
\begin{align} \label{eq:Jpc}
&2^N\,J^{\kappa[N]}\left(\ln \frac{x_L x^2 Q^2}{\mu^2}, \mu\right) =2^N\,\hat{J}^{\kappa[N]}\left(\ln \frac{x_L x^2 Q^2}{\mu^2}, \mu\right)\nonumber \\
&\qquad -\frac{N\,\overline{\Omega}_{1\kappa}}{\sqrt{x_L}x Q} \,2^{N-1}\,\hat{\mathcal{J}}^{\kappa[N-1]}\left(\ln \frac{x_L x^2 Q^2}{\mu^2}, \mu\right)\,,
\end{align}
where both $\hat{J}^{\kappa[N]}$ and $\hat{\cJ}^{\kappa[N-1]}$ are perturbative  $\overline{\mathrm{MS}}$ coefficients.
Next we define
\begin{align}
\hat J^{\kappa[N]} = \sum_{k = 0}^\infty \Big(\frac{\as(\mu)}{4\pi}\Big)^{k}\, \hat J^{\kappa[N]}_{k}	\,	,
\end{align}
as the perturbative series of the jet function, and analogously for $\hat{\cJ}^{\kappa[N-1]}$.
We use tree-level normalization $\hat{J}^{\kappa[N]}_0 = \hat{\cJ}^{\kappa[N]}_0 = 2^{-N}$, and explicit results for $\hat{J}^{\kappa[N]}_1$ can be found in~\refcite{Chen:2020vvp}.
Due to the presence of the $x^{-1}$ in the $\overline{\Omega}_{1\kappa}$ term, the logarithmic structure of the power correction contained in $\hat{\cJ}^{\kappa[N-1]}$ has the same evolution as $J^{\kappa[N-1]}$.%
\footnote{This reproduces the observation made by \refcite{HaoTalkSCET,Chen:2024nyc} that the spin of the anomalous dimension of the power suppressed term is decreased by $1$.}
Therefore, the single logarithm in  $\hat{\cJ}^{i[N-1]}_1$ is taken equal to that of  $\hat{J}^{i[N-1]}_1$, while the constant term can differ.
Computing the leading $u=1/2$ renormalon for $\hat J^{\kappa[N]}$ we find a result consistent with \eq{Jpc}.
An R scheme jet function which cancels the renormalon can be defined by
\begin{align}
&2^N \hat{J}_{\mathrm{R}}^{\kappa[N]}\left(\ln \frac{x_L Q^2}{\mu^2}, \mu\right) \equiv 2^N \hat{J}^{\kappa[N]}\left(\ln \frac{x_L Q^2}{\mu^2}, \mu\right) \\
& \qquad \qquad\qquad  -\sum_{n=1}^\infty
\frac{N\,R}{Q\sqrt{x_L}} \, d_{\kappa n}\bigl(\mu/R\bigr)
\Big(  \frac{\alpha_s(\mu)}{4\pi} \Big)^n\nonumber
\,,
\end{align}
where $d_{\kappa n}$ are identical to those in \eq{FORscheme}.

In \fig{transition}, we clearly see the importance of incorporating the $\Omega_{1q}$ nonperturbative corrections. Looking from the right, we first encounter the fixed order region shown in red, which is discussed in \fig{FO}.
We then hit the perturbative resummation region, shown in orange, where quarks and gluons radiate to give us the scaling predicted by the light-ray OPE. The radiating partons eventually lose enough energy and are confined to form a bound state of hadrons, which manifests itself in the turnover in the green region, now also observed in numerous experimental datasets across many collaborations~\cite{CMS:2024mlf,Mazzilli:2024ots,Tamis:2023guc,Fan2023}, and through string or clustering fragmentation~\cite{Andersson:1983ia,Marchesini:1991ch,Bahr:2008pv,Bahr:2008tf,Bellm:2019zci} in MC simulations as illustrated in the figure for \Pythia and \Herwig. Finally, we reach the free hadron region in blue, where correlations vanish.
The golden dashed line is a linear fit to the correlations in the free hadron region, $d \sigma^{[N]} \propto \theta_L$, corresponding to a uniform distribution in $\theta_L^2\sim x_L$.

\begin{figure}[t!]
\begin{center}
\includegraphics[scale=0.33]{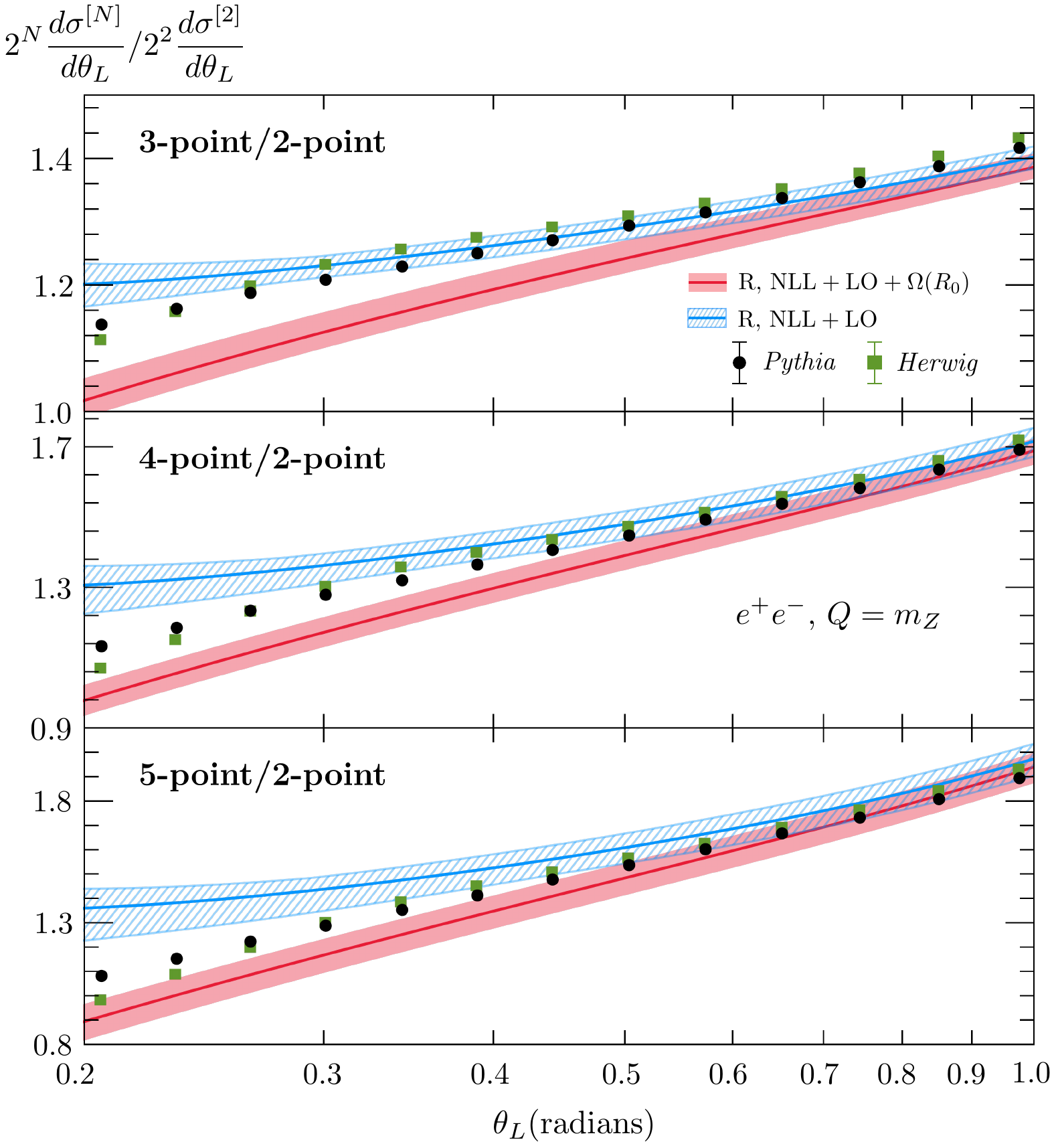}
\end{center}
\caption{The ratio of pENC to EEC probe the quantum scaling associated with higher spin light-ray operators. One can observe deviations from the linear behavior in the small angle region due to nonperturbative power corrections.}
\label{fig:ratio}
\end{figure}

\begin{figure}[t!]
\begin{center}
\includegraphics[scale=0.265]{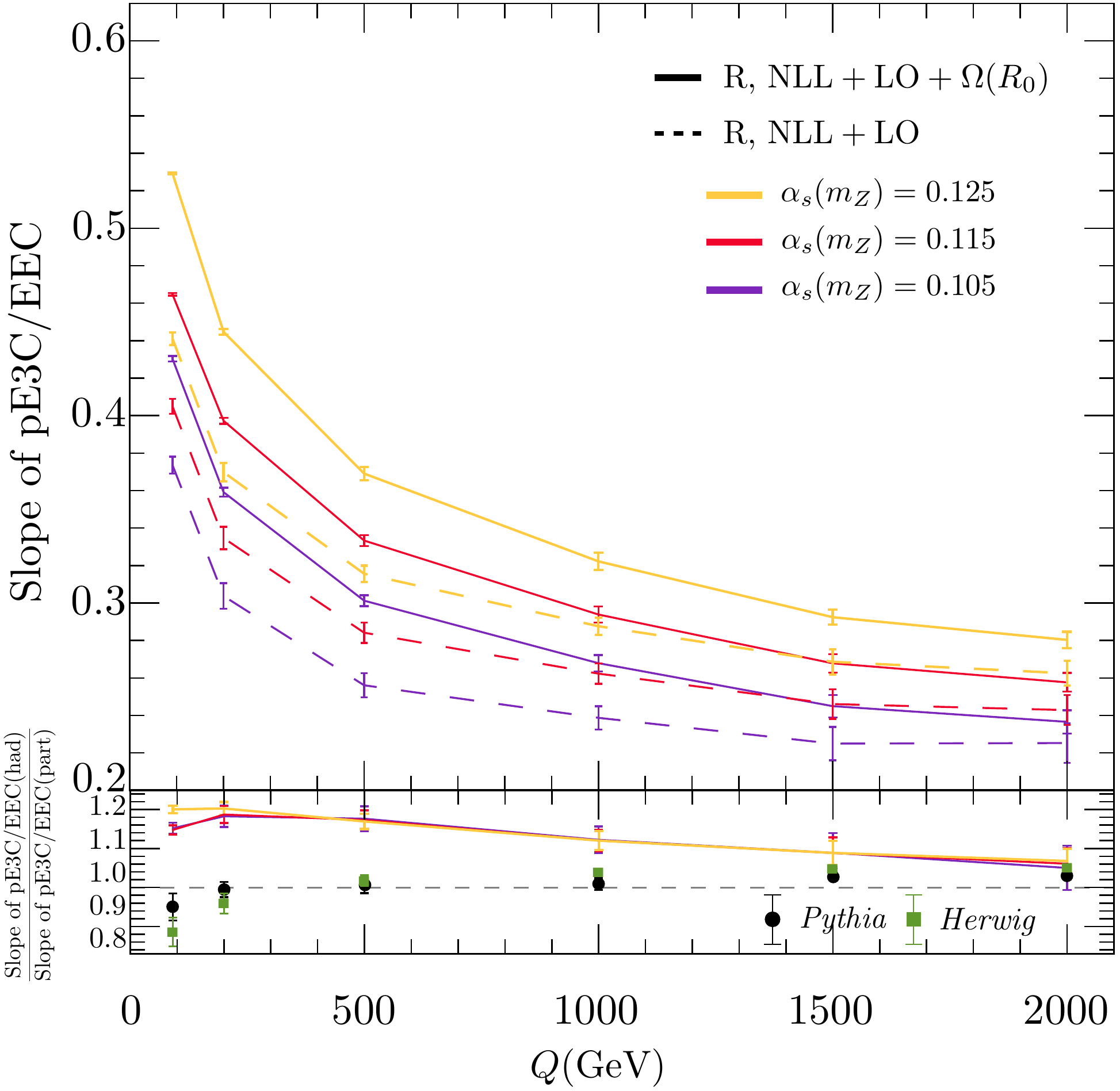}
\end{center}
\caption{We plot the ratio of pE3C to EEC at different $Q$ values at different values of $\alpha_s(m_Z)$ with and without nonperturbative power corrections.
}
\label{fig:slope}
\end{figure}

We see that incorporating nonperturbative power corrections in the R scheme significantly improves our description of the approach to the hadronization peak from the parton scaling region.
As we get closer to this peak, the nonperturbative effects no longer remain subleading.
Since the first term in \eq{pENCpc} scales as $1/(2^Nx_L)$, this occurs when $N\Omega_{1q} /(Q\sqrt{x_{L,\rm peak}}) \sim 1$ for the $N$-point correlator (up to anomalous scaling). For $\theta_{L,\rm{peak}}\sim \sqrt{x_{L,\rm{peak}}}$ this predicts a $N/2$ relation between the peak location for $N>2$ relative to $N=2$.
Fixing the constant using the EEC and  $\Omega_{1q}$ above, we find
for EEC $2\Omega_{1q} /(Q\sqrt{x_{L,{\rm peak}}}) \approx 2/3$, which then predicts
the vertical purple dotted lines in \fig{transition}, consistent with the peak locations for higher $N$.

\sectionPaper{Strong Coupling Determination}
By taking the ratio of pENCs to EEC, we eliminate their classical scaling and isolate the anomalous scaling proportional to $\alpha_s$,
making the ratio in the small angle region an ideal observable for cleanly extracting $\alpha_s$~\cite{Chen:2020vvp}. Due to the monotonicity of twist-2 spin-$(N+1)$ anomalous dimensions as a function of $N$, the slope of the ratio increases with $N$~\cite{Nachtmann:1973mr}. This method recently yielded the most precise jet substructure based $\alpha_s$ extraction at the LHC, with the CMS collaboration achieving an accuracy of 4\% with $\alpha_s(m_Z)=0.123^{+0.004}_{-0.005}$ by analyzing the ratio of pE3C to EEC for a range of jet $p_T$~\cite{CMS:2024mlf}. Generally, it was expected that nonperturbative contributions across different values of $N$ would cancel out in the ratio~\cite{Komiske:2022enw,Chen:2023zlx,CMS:2024mlf}. This expectation led CMS to use perturbative theoretical predictions~\cite{Chen:2023zlx}, combined with MC modeling of hadronization effects, in the CMS extraction. With our systematic prescription for removing renormalon ambiguities and our prediction for the relation between nonperturbative corrections for different $N$ values, we are well-positioned to assess the magnitude of these effects  in $\as$ determination using pENCs with our field theory based approach.

In \fig{ratio}, we compare the pENC/EEC ratio with and without nonperturbative power corrections at $Q=m_Z$ for $N=2-5$.
Although in the perturbative region shown one fits for $\alpha_s(m_Z)$ using the full functional form,
the impact of various parameters can be estimated from changes to the slope in a linear fit approximation, which becomes better further away from the transition region.
As the angle approaches the transition region, we observe that the MC simulations begin to deviate from the linear slope, which nonperturbative corrections capture well.

In \fig{slope}, we examine the impact of nonperturbative power corrections by plotting the slope of pE3C/EEC within the perturbative scaling region $3\times 10~{\rm{GeV}}/(Q/2)\leq \theta_L\leq 1$ to ensure separation from the transition region.
This visualization clearly shows the decreasing slope as we move to higher $Q$, serving as a clear signature of asymptotic freedom. Contrary to earlier observations, nonperturbative power corrections significantly impact the ratio.
In the lower panel of \fig{slope}, we plot the ratio of hadronic and partonic slopes, and see that the size of hadronization correction in our analytic calculation decreases from $\sim$20\% down to $\sim$5\% for increasing $Q$, and is much larger than the small effects observed in \Pythia and \Herwig.
This underscores the danger of solely relying on the difference between hadronic and partonic MC to model hadronization corrections in perturbative calculations.

Comparing the dashed and solid lines in \fig{slope} shows a degeneracy between including nonperturbative power corrections and increasing $\alpha_s(m_Z)$ values. This illustrates the importance of controlling nonperturbative corrections to achieve a precise fit for $\alpha_s$.  For example, at $Q=1000$ GeV including nonperturbative corrections is a $\sim$10\% effect in the direction of decreasing $\alpha_s$,
which is significantly larger than the small~\cite{Komiske:2022enw,Chen:2023zlx} $\sim 0 -3\%$ effect estimated by MCs in the CMS analysis~\cite{CMS:2024mlf}.

\sectionPaper{Conclusion}
We performed a model-independent assessment of nonperturbative corrections on projected $N$-point energy correlators, and
derived a relation between these corrections. Our predictions in a renormalon-free scheme improve hadron-level results for both large and small angle regions.
This analysis
provides a strong motivation for revisiting the $\as$ determination using our model-independent predictions for nonperturbative corrections.
In particular, we found that MCs significantly underestimate the size of the nonperturbative effects.
This is particularly pertinent to upcoming precision analyses of LEP data and jet measurements at
existing and future colliders, including the LHC, RHIC, and EIC.

\begin{acknowledgments}
\headingAcknowledgments
We thank Hao Chen, Philip Harris, Andre Hoang, Ian Moult, Xiaoyuan Zhang, and Hua Xing Zhu for helpful discussions. We thank the Erwin Schr\"{o}dinger Institute for hospitality while parts of this work were performed. K.L., I.S., and Z.S were supported by the U.S. Department of Energy, Office of Science, Office of Nuclear Physics from DE-SC0011090. I.S. was also supported in part by the Simons Foundation through the Investigator grant 327942. Z.S. was also supported by a fellowship from the MIT Department of Physics.

\end{acknowledgments}

\bibliography{refs}

\begin{thebibliography}{88}%
\makeatletter
\providecommand \@ifxundefined [1]{%
 \@ifx{#1\undefined}
}%
\providecommand \@ifnum [1]{%
 \ifnum #1\expandafter \@firstoftwo
 \else \expandafter \@secondoftwo
 \fi
}%
\providecommand \@ifx [1]{%
 \ifx #1\expandafter \@firstoftwo
 \else \expandafter \@secondoftwo
 \fi
}%
\providecommand \natexlab [1]{#1}%
\providecommand \enquote  [1]{``#1''}%
\providecommand \bibnamefont  [1]{#1}%
\providecommand \bibfnamefont [1]{#1}%
\providecommand \citenamefont [1]{#1}%
\providecommand \href@noop [0]{\@secondoftwo}%
\providecommand \href [0]{\begingroup \@sanitize@url \@href}%
\providecommand \@href[1]{\@@startlink{#1}\@@href}%
\providecommand \@@href[1]{\endgroup#1\@@endlink}%
\providecommand \@sanitize@url [0]{\catcode `\\12\catcode `\$12\catcode
  `\&12\catcode `\#12\catcode `\^12\catcode `\_12\catcode `\%12\relax}%
\providecommand \@@startlink[1]{}%
\providecommand \@@endlink[0]{}%
\providecommand \url  [0]{\begingroup\@sanitize@url \@url }%
\providecommand \@url [1]{\endgroup\@href {#1}{\urlprefix }}%
\providecommand \urlprefix  [0]{URL }%
\providecommand \Eprint [0]{\href }%
\providecommand \doibase [0]{https://doi.org/}%
\providecommand \selectlanguage [0]{\@gobble}%
\providecommand \bibinfo  [0]{\@secondoftwo}%
\providecommand \bibfield  [0]{\@secondoftwo}%
\providecommand \translation [1]{[#1]}%
\providecommand \BibitemOpen [0]{}%
\providecommand \bibitemStop [0]{}%
\providecommand \bibitemNoStop [0]{.\EOS\space}%
\providecommand \EOS [0]{\spacefactor3000\relax}%
\providecommand \BibitemShut  [1]{\csname bibitem#1\endcsname}%
\let\auto@bib@innerbib\@empty
\bibitem [{\citenamefont {Sveshnikov}\ and\ \citenamefont
  {Tkachov}(1996)}]{Sveshnikov:1995vi}%
  \BibitemOpen
  \bibfield  {author} {\bibinfo {author} {\bibfnamefont {N.}~\bibnamefont
  {Sveshnikov}}\ and\ \bibinfo {author} {\bibfnamefont {F.}~\bibnamefont
  {Tkachov}},\ }\bibfield  {title} {\bibinfo {title} {{Jets and quantum field
  theory}},\ }\href {https://doi.org/10.1016/0370-2693(96)00558-8} {\bibfield
  {journal} {\bibinfo  {journal} {Phys. Lett. B}\ }\textbf {\bibinfo {volume}
  {382}},\ \bibinfo {pages} {403} (\bibinfo {year} {1996})},\ \Eprint
  {https://arxiv.org/abs/hep-ph/9512370} {arXiv:hep-ph/9512370} \BibitemShut
  {NoStop}%
\bibitem [{\citenamefont {Tkachov}(1997)}]{Tkachov:1995kk}%
  \BibitemOpen
  \bibfield  {author} {\bibinfo {author} {\bibfnamefont {F.~V.}\ \bibnamefont
  {Tkachov}},\ }\bibfield  {title} {\bibinfo {title} {{Measuring multi - jet
  structure of hadronic energy flow or What is a jet?}},\ }\href
  {https://doi.org/10.1142/S0217751X97002899} {\bibfield  {journal} {\bibinfo
  {journal} {Int. J. Mod. Phys. A}\ }\textbf {\bibinfo {volume} {12}},\
  \bibinfo {pages} {5411} (\bibinfo {year} {1997})},\ \Eprint
  {https://arxiv.org/abs/hep-ph/9601308} {arXiv:hep-ph/9601308} \BibitemShut
  {NoStop}%
\bibitem [{\citenamefont {Korchemsky}\ and\ \citenamefont
  {Sterman}(1999)}]{Korchemsky:1999kt}%
  \BibitemOpen
  \bibfield  {author} {\bibinfo {author} {\bibfnamefont {G.~P.}\ \bibnamefont
  {Korchemsky}}\ and\ \bibinfo {author} {\bibfnamefont {G.~F.}\ \bibnamefont
  {Sterman}},\ }\bibfield  {title} {\bibinfo {title} {{Power corrections to
  event shapes and factorization}},\ }\href
  {https://doi.org/10.1016/S0550-3213(99)00308-9} {\bibfield  {journal}
  {\bibinfo  {journal} {Nucl. Phys. B}\ }\textbf {\bibinfo {volume} {555}},\
  \bibinfo {pages} {335} (\bibinfo {year} {1999})},\ \Eprint
  {https://arxiv.org/abs/hep-ph/9902341} {arXiv:hep-ph/9902341} \BibitemShut
  {NoStop}%
\bibitem [{\citenamefont {Bauer}\ \emph {et~al.}(2008)\citenamefont {Bauer},
  \citenamefont {Fleming}, \citenamefont {Lee},\ and\ \citenamefont
  {Sterman}}]{Bauer:2008dt}%
  \BibitemOpen
  \bibfield  {author} {\bibinfo {author} {\bibfnamefont {C.~W.}\ \bibnamefont
  {Bauer}}, \bibinfo {author} {\bibfnamefont {S.~P.}\ \bibnamefont {Fleming}},
  \bibinfo {author} {\bibfnamefont {C.}~\bibnamefont {Lee}},\ and\ \bibinfo
  {author} {\bibfnamefont {G.~F.}\ \bibnamefont {Sterman}},\ }\bibfield
  {title} {\bibinfo {title} {{Factorization of e+e- Event Shape Distributions
  with Hadronic Final States in Soft Collinear Effective Theory}},\ }\href
  {https://doi.org/10.1103/PhysRevD.78.034027} {\bibfield  {journal} {\bibinfo
  {journal} {Phys. Rev. D}\ }\textbf {\bibinfo {volume} {78}},\ \bibinfo
  {pages} {034027} (\bibinfo {year} {2008})},\ \Eprint
  {https://arxiv.org/abs/0801.4569} {arXiv:0801.4569 [hep-ph]} \BibitemShut
  {NoStop}%
\bibitem [{\citenamefont {Hofman}\ and\ \citenamefont
  {Maldacena}(2008)}]{Hofman:2008ar}%
  \BibitemOpen
  \bibfield  {author} {\bibinfo {author} {\bibfnamefont {D.~M.}\ \bibnamefont
  {Hofman}}\ and\ \bibinfo {author} {\bibfnamefont {J.}~\bibnamefont
  {Maldacena}},\ }\bibfield  {title} {\bibinfo {title} {{Conformal collider
  physics: Energy and charge correlations}},\ }\href
  {https://doi.org/10.1088/1126-6708/2008/05/012} {\bibfield  {journal}
  {\bibinfo  {journal} {JHEP}\ }\textbf {\bibinfo {volume} {05}},\ \bibinfo
  {pages} {012}},\ \Eprint {https://arxiv.org/abs/0803.1467} {arXiv:0803.1467
  [hep-th]} \BibitemShut {NoStop}%
\bibitem [{\citenamefont {Belitsky}\ \emph
  {et~al.}(2014{\natexlab{a}})\citenamefont {Belitsky}, \citenamefont
  {Hohenegger}, \citenamefont {Korchemsky}, \citenamefont {Sokatchev},\ and\
  \citenamefont {Zhiboedov}}]{Belitsky:2013xxa}%
  \BibitemOpen
  \bibfield  {author} {\bibinfo {author} {\bibfnamefont {A.}~\bibnamefont
  {Belitsky}}, \bibinfo {author} {\bibfnamefont {S.}~\bibnamefont
  {Hohenegger}}, \bibinfo {author} {\bibfnamefont {G.}~\bibnamefont
  {Korchemsky}}, \bibinfo {author} {\bibfnamefont {E.}~\bibnamefont
  {Sokatchev}},\ and\ \bibinfo {author} {\bibfnamefont {A.}~\bibnamefont
  {Zhiboedov}},\ }\bibfield  {title} {\bibinfo {title} {{From correlation
  functions to event shapes}},\ }\href
  {https://doi.org/10.1016/j.nuclphysb.2014.04.020} {\bibfield  {journal}
  {\bibinfo  {journal} {Nucl. Phys. B}\ }\textbf {\bibinfo {volume} {884}},\
  \bibinfo {pages} {305} (\bibinfo {year} {2014}{\natexlab{a}})},\ \Eprint
  {https://arxiv.org/abs/1309.0769} {arXiv:1309.0769 [hep-th]} \BibitemShut
  {NoStop}%
\bibitem [{\citenamefont {Belitsky}\ \emph
  {et~al.}(2014{\natexlab{b}})\citenamefont {Belitsky}, \citenamefont
  {Hohenegger}, \citenamefont {Korchemsky}, \citenamefont {Sokatchev},\ and\
  \citenamefont {Zhiboedov}}]{Belitsky:2013bja}%
  \BibitemOpen
  \bibfield  {author} {\bibinfo {author} {\bibfnamefont {A.}~\bibnamefont
  {Belitsky}}, \bibinfo {author} {\bibfnamefont {S.}~\bibnamefont
  {Hohenegger}}, \bibinfo {author} {\bibfnamefont {G.}~\bibnamefont
  {Korchemsky}}, \bibinfo {author} {\bibfnamefont {E.}~\bibnamefont
  {Sokatchev}},\ and\ \bibinfo {author} {\bibfnamefont {A.}~\bibnamefont
  {Zhiboedov}},\ }\bibfield  {title} {\bibinfo {title} {{Event shapes in
  $\mathcal{N} = 4$ super-Yang-Mills theory}},\ }\href
  {https://doi.org/10.1016/j.nuclphysb.2014.04.019} {\bibfield  {journal}
  {\bibinfo  {journal} {Nucl. Phys. B}\ }\textbf {\bibinfo {volume} {884}},\
  \bibinfo {pages} {206} (\bibinfo {year} {2014}{\natexlab{b}})},\ \Eprint
  {https://arxiv.org/abs/1309.1424} {arXiv:1309.1424 [hep-th]} \BibitemShut
  {NoStop}%
\bibitem [{\citenamefont {Kravchuk}\ and\ \citenamefont
  {Simmons-Duffin}(2018)}]{Kravchuk:2018htv}%
  \BibitemOpen
  \bibfield  {author} {\bibinfo {author} {\bibfnamefont {P.}~\bibnamefont
  {Kravchuk}}\ and\ \bibinfo {author} {\bibfnamefont {D.}~\bibnamefont
  {Simmons-Duffin}},\ }\bibfield  {title} {\bibinfo {title} {{Light-ray
  operators in conformal field theory}},\ }\href
  {https://doi.org/10.1007/JHEP11(2018)102} {\bibfield  {journal} {\bibinfo
  {journal} {JHEP}\ }\textbf {\bibinfo {volume} {11}},\ \bibinfo {pages}
  {102}},\ \Eprint {https://arxiv.org/abs/1805.00098} {arXiv:1805.00098
  [hep-th]} \BibitemShut {NoStop}%
\bibitem [{\citenamefont {Basham}\ \emph
  {et~al.}(1979{\natexlab{a}})\citenamefont {Basham}, \citenamefont {Brown},
  \citenamefont {Ellis},\ and\ \citenamefont {Love}}]{Basham:1979gh}%
  \BibitemOpen
  \bibfield  {author} {\bibinfo {author} {\bibfnamefont {C.~L.}\ \bibnamefont
  {Basham}}, \bibinfo {author} {\bibfnamefont {L.~S.}\ \bibnamefont {Brown}},
  \bibinfo {author} {\bibfnamefont {S.~D.}\ \bibnamefont {Ellis}},\ and\
  \bibinfo {author} {\bibfnamefont {S.~T.}\ \bibnamefont {Love}},\ }\bibfield
  {title} {\bibinfo {title} {{Energy Correlations in Perturbative Quantum
  Chromodynamics: A Conjecture for All Orders}},\ }\href
  {https://doi.org/10.1016/0370-2693(79)90601-4} {\bibfield  {journal}
  {\bibinfo  {journal} {Phys. Lett. B}\ }\textbf {\bibinfo {volume} {85}},\
  \bibinfo {pages} {297} (\bibinfo {year} {1979}{\natexlab{a}})}\BibitemShut
  {NoStop}%
\bibitem [{\citenamefont {Basham}\ \emph
  {et~al.}(1979{\natexlab{b}})\citenamefont {Basham}, \citenamefont {Brown},
  \citenamefont {Ellis},\ and\ \citenamefont {Love}}]{Basham:1978zq}%
  \BibitemOpen
  \bibfield  {author} {\bibinfo {author} {\bibfnamefont {C.}~\bibnamefont
  {Basham}}, \bibinfo {author} {\bibfnamefont {L.}~\bibnamefont {Brown}},
  \bibinfo {author} {\bibfnamefont {S.}~\bibnamefont {Ellis}},\ and\ \bibinfo
  {author} {\bibfnamefont {S.}~\bibnamefont {Love}},\ }\bibfield  {title}
  {\bibinfo {title} {{Energy Correlations in electron-Positron Annihilation in
  Quantum Chromodynamics: Asymptotically Free Perturbation Theory}},\ }\href
  {https://doi.org/10.1103/PhysRevD.19.2018} {\bibfield  {journal} {\bibinfo
  {journal} {Phys. Rev. D}\ }\textbf {\bibinfo {volume} {19}},\ \bibinfo
  {pages} {2018} (\bibinfo {year} {1979}{\natexlab{b}})}\BibitemShut {NoStop}%
\bibitem [{\citenamefont {Basham}\ \emph
  {et~al.}(1978{\natexlab{a}})\citenamefont {Basham}, \citenamefont {Brown},
  \citenamefont {Ellis},\ and\ \citenamefont {Love}}]{Basham:1978bw}%
  \BibitemOpen
  \bibfield  {author} {\bibinfo {author} {\bibfnamefont {C.}~\bibnamefont
  {Basham}}, \bibinfo {author} {\bibfnamefont {L.~S.}\ \bibnamefont {Brown}},
  \bibinfo {author} {\bibfnamefont {S.~D.}\ \bibnamefont {Ellis}},\ and\
  \bibinfo {author} {\bibfnamefont {S.~T.}\ \bibnamefont {Love}},\ }\bibfield
  {title} {\bibinfo {title} {{Energy Correlations in electron - Positron
  Annihilation: Testing QCD}},\ }\href
  {https://doi.org/10.1103/PhysRevLett.41.1585} {\bibfield  {journal} {\bibinfo
   {journal} {Phys. Rev. Lett.}\ }\textbf {\bibinfo {volume} {41}},\ \bibinfo
  {pages} {1585} (\bibinfo {year} {1978}{\natexlab{a}})}\BibitemShut {NoStop}%
\bibitem [{\citenamefont {Basham}\ \emph
  {et~al.}(1978{\natexlab{b}})\citenamefont {Basham}, \citenamefont {Brown},
  \citenamefont {Ellis},\ and\ \citenamefont {Love}}]{Basham:1977iq}%
  \BibitemOpen
  \bibfield  {author} {\bibinfo {author} {\bibfnamefont {C.~L.}\ \bibnamefont
  {Basham}}, \bibinfo {author} {\bibfnamefont {L.~S.}\ \bibnamefont {Brown}},
  \bibinfo {author} {\bibfnamefont {S.~D.}\ \bibnamefont {Ellis}},\ and\
  \bibinfo {author} {\bibfnamefont {S.~T.}\ \bibnamefont {Love}},\ }\bibfield
  {title} {\bibinfo {title} {{Electron - Positron Annihilation Energy Pattern
  in Quantum Chromodynamics: Asymptotically Free Perturbation Theory}},\ }\href
  {https://doi.org/10.1103/PhysRevD.17.2298} {\bibfield  {journal} {\bibinfo
  {journal} {Phys. Rev. D}\ }\textbf {\bibinfo {volume} {17}},\ \bibinfo
  {pages} {2298} (\bibinfo {year} {1978}{\natexlab{b}})}\BibitemShut {NoStop}%
\bibitem [{\citenamefont {Dixon}\ \emph {et~al.}(2019)\citenamefont {Dixon},
  \citenamefont {Moult},\ and\ \citenamefont {Zhu}}]{Dixon:2019uzg}%
  \BibitemOpen
  \bibfield  {author} {\bibinfo {author} {\bibfnamefont {L.~J.}\ \bibnamefont
  {Dixon}}, \bibinfo {author} {\bibfnamefont {I.}~\bibnamefont {Moult}},\ and\
  \bibinfo {author} {\bibfnamefont {H.~X.}\ \bibnamefont {Zhu}},\ }\bibfield
  {title} {\bibinfo {title} {{Collinear limit of the energy-energy
  correlator}},\ }\href {https://doi.org/10.1103/PhysRevD.100.014009}
  {\bibfield  {journal} {\bibinfo  {journal} {Phys. Rev. D}\ }\textbf {\bibinfo
  {volume} {100}},\ \bibinfo {pages} {014009} (\bibinfo {year} {2019})},\
  \Eprint {https://arxiv.org/abs/1905.01310} {arXiv:1905.01310 [hep-ph]}
  \BibitemShut {NoStop}%
\bibitem [{\citenamefont {Chen}\ \emph {et~al.}(2020)\citenamefont {Chen},
  \citenamefont {Moult}, \citenamefont {Zhang},\ and\ \citenamefont
  {Zhu}}]{Chen:2020vvp}%
  \BibitemOpen
  \bibfield  {author} {\bibinfo {author} {\bibfnamefont {H.}~\bibnamefont
  {Chen}}, \bibinfo {author} {\bibfnamefont {I.}~\bibnamefont {Moult}},
  \bibinfo {author} {\bibfnamefont {X.}~\bibnamefont {Zhang}},\ and\ \bibinfo
  {author} {\bibfnamefont {H.~X.}\ \bibnamefont {Zhu}},\ }\bibfield  {title}
  {\bibinfo {title} {{Rethinking jets with energy correlators: Tracks,
  resummation, and analytic continuation}},\ }\href
  {https://doi.org/10.1103/PhysRevD.102.054012} {\bibfield  {journal} {\bibinfo
   {journal} {Phys. Rev. D}\ }\textbf {\bibinfo {volume} {102}},\ \bibinfo
  {pages} {054012} (\bibinfo {year} {2020})},\ \Eprint
  {https://arxiv.org/abs/2004.11381} {arXiv:2004.11381 [hep-ph]} \BibitemShut
  {NoStop}%
\bibitem [{\citenamefont {Lee}\ \emph {et~al.}(2022)\citenamefont {Lee},
  \citenamefont {Me\c{c}aj},\ and\ \citenamefont {Moult}}]{Lee:2022ige}%
  \BibitemOpen
  \bibfield  {author} {\bibinfo {author} {\bibfnamefont {K.}~\bibnamefont
  {Lee}}, \bibinfo {author} {\bibfnamefont {B.}~\bibnamefont {Me\c{c}aj}},\
  and\ \bibinfo {author} {\bibfnamefont {I.}~\bibnamefont {Moult}},\ }\bibfield
   {title} {\bibinfo {title} {{Conformal Colliders Meet the LHC}},\ }\href@noop
  {} {\  (\bibinfo {year} {2022})},\ \Eprint {https://arxiv.org/abs/2205.03414}
  {arXiv:2205.03414 [hep-ph]} \BibitemShut {NoStop}%
\bibitem [{\citenamefont {Barata}\ \emph {et~al.}(2023)\citenamefont {Barata},
  \citenamefont {Caucal}, \citenamefont {Soto-Ontoso},\ and\ \citenamefont
  {Szafron}}]{Barata:2023bhh}%
  \BibitemOpen
  \bibfield  {author} {\bibinfo {author} {\bibfnamefont {J.~a.}\ \bibnamefont
  {Barata}}, \bibinfo {author} {\bibfnamefont {P.}~\bibnamefont {Caucal}},
  \bibinfo {author} {\bibfnamefont {A.}~\bibnamefont {Soto-Ontoso}},\ and\
  \bibinfo {author} {\bibfnamefont {R.}~\bibnamefont {Szafron}},\ }\bibfield
  {title} {\bibinfo {title} {{Advancing the understanding of energy-energy
  correlators in heavy-ion collisions}},\ }\href@noop {} {\  (\bibinfo {year}
  {2023})},\ \Eprint {https://arxiv.org/abs/2312.12527} {arXiv:2312.12527
  [hep-ph]} \BibitemShut {NoStop}%
\bibitem [{\citenamefont {Komiske}\ \emph {et~al.}(2023)\citenamefont
  {Komiske}, \citenamefont {Moult}, \citenamefont {Thaler},\ and\ \citenamefont
  {Zhu}}]{Komiske:2022enw}%
  \BibitemOpen
  \bibfield  {author} {\bibinfo {author} {\bibfnamefont {P.~T.}\ \bibnamefont
  {Komiske}}, \bibinfo {author} {\bibfnamefont {I.}~\bibnamefont {Moult}},
  \bibinfo {author} {\bibfnamefont {J.}~\bibnamefont {Thaler}},\ and\ \bibinfo
  {author} {\bibfnamefont {H.~X.}\ \bibnamefont {Zhu}},\ }\bibfield  {title}
  {\bibinfo {title} {{Analyzing N-Point Energy Correlators inside Jets with CMS
  Open Data}},\ }\href {https://doi.org/10.1103/PhysRevLett.130.051901}
  {\bibfield  {journal} {\bibinfo  {journal} {Phys. Rev. Lett.}\ }\textbf
  {\bibinfo {volume} {130}},\ \bibinfo {pages} {051901} (\bibinfo {year}
  {2023})},\ \Eprint {https://arxiv.org/abs/2201.07800} {arXiv:2201.07800
  [hep-ph]} \BibitemShut {NoStop}%
\bibitem [{\citenamefont {Holguin}\ \emph
  {et~al.}(2023{\natexlab{a}})\citenamefont {Holguin}, \citenamefont {Moult},
  \citenamefont {Pathak},\ and\ \citenamefont {Procura}}]{Holguin:2022epo}%
  \BibitemOpen
  \bibfield  {author} {\bibinfo {author} {\bibfnamefont {J.}~\bibnamefont
  {Holguin}}, \bibinfo {author} {\bibfnamefont {I.}~\bibnamefont {Moult}},
  \bibinfo {author} {\bibfnamefont {A.}~\bibnamefont {Pathak}},\ and\ \bibinfo
  {author} {\bibfnamefont {M.}~\bibnamefont {Procura}},\ }\bibfield  {title}
  {\bibinfo {title} {{New paradigm for precision top physics: Weighing the top
  with energy correlators}},\ }\href
  {https://doi.org/10.1103/PhysRevD.107.114002} {\bibfield  {journal} {\bibinfo
   {journal} {Phys. Rev. D}\ }\textbf {\bibinfo {volume} {107}},\ \bibinfo
  {pages} {114002} (\bibinfo {year} {2023}{\natexlab{a}})},\ \Eprint
  {https://arxiv.org/abs/2201.08393} {arXiv:2201.08393 [hep-ph]} \BibitemShut
  {NoStop}%
\bibitem [{\citenamefont {Liu}\ and\ \citenamefont {Zhu}(2023)}]{Liu:2022wop}%
  \BibitemOpen
  \bibfield  {author} {\bibinfo {author} {\bibfnamefont {X.}~\bibnamefont
  {Liu}}\ and\ \bibinfo {author} {\bibfnamefont {H.~X.}\ \bibnamefont {Zhu}},\
  }\bibfield  {title} {\bibinfo {title} {{Nucleon Energy Correlators}},\ }\href
  {https://doi.org/10.1103/PhysRevLett.130.091901} {\bibfield  {journal}
  {\bibinfo  {journal} {Phys. Rev. Lett.}\ }\textbf {\bibinfo {volume} {130}},\
  \bibinfo {pages} {091901} (\bibinfo {year} {2023})},\ \Eprint
  {https://arxiv.org/abs/2209.02080} {arXiv:2209.02080 [hep-ph]} \BibitemShut
  {NoStop}%
\bibitem [{\citenamefont {Liu}\ \emph {et~al.}(2023)\citenamefont {Liu},
  \citenamefont {Liu}, \citenamefont {Pan}, \citenamefont {Yuan},\ and\
  \citenamefont {Zhu}}]{Liu:2023aqb}%
  \BibitemOpen
  \bibfield  {author} {\bibinfo {author} {\bibfnamefont {H.-Y.}\ \bibnamefont
  {Liu}}, \bibinfo {author} {\bibfnamefont {X.}~\bibnamefont {Liu}}, \bibinfo
  {author} {\bibfnamefont {J.-C.}\ \bibnamefont {Pan}}, \bibinfo {author}
  {\bibfnamefont {F.}~\bibnamefont {Yuan}},\ and\ \bibinfo {author}
  {\bibfnamefont {H.~X.}\ \bibnamefont {Zhu}},\ }\bibfield  {title} {\bibinfo
  {title} {{Nucleon Energy Correlators for the Color Glass Condensate}},\
  }\href {https://doi.org/10.1103/PhysRevLett.130.181901} {\bibfield  {journal}
  {\bibinfo  {journal} {Phys. Rev. Lett.}\ }\textbf {\bibinfo {volume} {130}},\
  \bibinfo {pages} {181901} (\bibinfo {year} {2023})},\ \Eprint
  {https://arxiv.org/abs/2301.01788} {arXiv:2301.01788 [hep-ph]} \BibitemShut
  {NoStop}%
\bibitem [{\citenamefont {Cao}\ \emph {et~al.}(2023)\citenamefont {Cao},
  \citenamefont {Liu},\ and\ \citenamefont {Zhu}}]{Cao:2023oef}%
  \BibitemOpen
  \bibfield  {author} {\bibinfo {author} {\bibfnamefont {H.}~\bibnamefont
  {Cao}}, \bibinfo {author} {\bibfnamefont {X.}~\bibnamefont {Liu}},\ and\
  \bibinfo {author} {\bibfnamefont {H.~X.}\ \bibnamefont {Zhu}},\ }\bibfield
  {title} {\bibinfo {title} {{Toward precision measurements of nucleon energy
  correlators in lepton-nucleon collisions}},\ }\href
  {https://doi.org/10.1103/PhysRevD.107.114008} {\bibfield  {journal} {\bibinfo
   {journal} {Phys. Rev. D}\ }\textbf {\bibinfo {volume} {107}},\ \bibinfo
  {pages} {114008} (\bibinfo {year} {2023})},\ \Eprint
  {https://arxiv.org/abs/2303.01530} {arXiv:2303.01530 [hep-ph]} \BibitemShut
  {NoStop}%
\bibitem [{\citenamefont {Devereaux}\ \emph {et~al.}(2023)\citenamefont
  {Devereaux}, \citenamefont {Fan}, \citenamefont {Ke}, \citenamefont {Lee},\
  and\ \citenamefont {Moult}}]{Devereaux:2023vjz}%
  \BibitemOpen
  \bibfield  {author} {\bibinfo {author} {\bibfnamefont {K.}~\bibnamefont
  {Devereaux}}, \bibinfo {author} {\bibfnamefont {W.}~\bibnamefont {Fan}},
  \bibinfo {author} {\bibfnamefont {W.}~\bibnamefont {Ke}}, \bibinfo {author}
  {\bibfnamefont {K.}~\bibnamefont {Lee}},\ and\ \bibinfo {author}
  {\bibfnamefont {I.}~\bibnamefont {Moult}},\ }\bibfield  {title} {\bibinfo
  {title} {{Imaging Cold Nuclear Matter with Energy Correlators}},\ }\href@noop
  {} {\  (\bibinfo {year} {2023})},\ \Eprint {https://arxiv.org/abs/2303.08143}
  {arXiv:2303.08143 [hep-ph]} \BibitemShut {NoStop}%
\bibitem [{\citenamefont {Andres}\ \emph
  {et~al.}(2023{\natexlab{a}})\citenamefont {Andres}, \citenamefont
  {Dominguez}, \citenamefont {Kunnawalkam~Elayavalli}, \citenamefont {Holguin},
  \citenamefont {Marquet},\ and\ \citenamefont {Moult}}]{Andres:2022ovj}%
  \BibitemOpen
  \bibfield  {author} {\bibinfo {author} {\bibfnamefont {C.}~\bibnamefont
  {Andres}}, \bibinfo {author} {\bibfnamefont {F.}~\bibnamefont {Dominguez}},
  \bibinfo {author} {\bibfnamefont {R.}~\bibnamefont {Kunnawalkam~Elayavalli}},
  \bibinfo {author} {\bibfnamefont {J.}~\bibnamefont {Holguin}}, \bibinfo
  {author} {\bibfnamefont {C.}~\bibnamefont {Marquet}},\ and\ \bibinfo {author}
  {\bibfnamefont {I.}~\bibnamefont {Moult}},\ }\bibfield  {title} {\bibinfo
  {title} {{Resolving the Scales of the Quark-Gluon Plasma with Energy
  Correlators}},\ }\href {https://doi.org/10.1103/PhysRevLett.130.262301}
  {\bibfield  {journal} {\bibinfo  {journal} {Phys. Rev. Lett.}\ }\textbf
  {\bibinfo {volume} {130}},\ \bibinfo {pages} {262301} (\bibinfo {year}
  {2023}{\natexlab{a}})},\ \Eprint {https://arxiv.org/abs/2209.11236}
  {arXiv:2209.11236 [hep-ph]} \BibitemShut {NoStop}%
\bibitem [{\citenamefont {Andres}\ \emph
  {et~al.}(2023{\natexlab{b}})\citenamefont {Andres}, \citenamefont
  {Dominguez}, \citenamefont {Holguin}, \citenamefont {Marquet},\ and\
  \citenamefont {Moult}}]{Andres:2023xwr}%
  \BibitemOpen
  \bibfield  {author} {\bibinfo {author} {\bibfnamefont {C.}~\bibnamefont
  {Andres}}, \bibinfo {author} {\bibfnamefont {F.}~\bibnamefont {Dominguez}},
  \bibinfo {author} {\bibfnamefont {J.}~\bibnamefont {Holguin}}, \bibinfo
  {author} {\bibfnamefont {C.}~\bibnamefont {Marquet}},\ and\ \bibinfo {author}
  {\bibfnamefont {I.}~\bibnamefont {Moult}},\ }\bibfield  {title} {\bibinfo
  {title} {{A coherent view of the quark-gluon plasma from energy
  correlators}},\ }\href {https://doi.org/10.1007/JHEP09(2023)088} {\bibfield
  {journal} {\bibinfo  {journal} {JHEP}\ }\textbf {\bibinfo {volume} {09}},\
  \bibinfo {pages} {088}},\ \Eprint {https://arxiv.org/abs/2303.03413}
  {arXiv:2303.03413 [hep-ph]} \BibitemShut {NoStop}%
\bibitem [{\citenamefont {Craft}\ \emph {et~al.}(2022)\citenamefont {Craft},
  \citenamefont {Lee}, \citenamefont {Me\c{c}aj},\ and\ \citenamefont
  {Moult}}]{Craft:2022kdo}%
  \BibitemOpen
  \bibfield  {author} {\bibinfo {author} {\bibfnamefont {E.}~\bibnamefont
  {Craft}}, \bibinfo {author} {\bibfnamefont {K.}~\bibnamefont {Lee}}, \bibinfo
  {author} {\bibfnamefont {B.}~\bibnamefont {Me\c{c}aj}},\ and\ \bibinfo
  {author} {\bibfnamefont {I.}~\bibnamefont {Moult}},\ }\bibfield  {title}
  {\bibinfo {title} {{Beautiful and Charming Energy Correlators}},\ }\href@noop
  {} {\  (\bibinfo {year} {2022})},\ \Eprint {https://arxiv.org/abs/2210.09311}
  {arXiv:2210.09311 [hep-ph]} \BibitemShut {NoStop}%
\bibitem [{\citenamefont {Lee}\ and\ \citenamefont
  {Moult}(2023)}]{Lee:2023npz}%
  \BibitemOpen
  \bibfield  {author} {\bibinfo {author} {\bibfnamefont {K.}~\bibnamefont
  {Lee}}\ and\ \bibinfo {author} {\bibfnamefont {I.}~\bibnamefont {Moult}},\
  }\bibfield  {title} {\bibinfo {title} {{Energy Correlators Taking Charge}},\
  }\href@noop {} {\  (\bibinfo {year} {2023})},\ \Eprint
  {https://arxiv.org/abs/2308.00746} {arXiv:2308.00746 [hep-ph]} \BibitemShut
  {NoStop}%
\bibitem [{\citenamefont {Holguin}\ \emph
  {et~al.}(2023{\natexlab{b}})\citenamefont {Holguin}, \citenamefont {Moult},
  \citenamefont {Pathak}, \citenamefont {Procura}, \citenamefont
  {Sch\"ofbeck},\ and\ \citenamefont {Schwarz}}]{Holguin:2023bjf}%
  \BibitemOpen
  \bibfield  {author} {\bibinfo {author} {\bibfnamefont {J.}~\bibnamefont
  {Holguin}}, \bibinfo {author} {\bibfnamefont {I.}~\bibnamefont {Moult}},
  \bibinfo {author} {\bibfnamefont {A.}~\bibnamefont {Pathak}}, \bibinfo
  {author} {\bibfnamefont {M.}~\bibnamefont {Procura}}, \bibinfo {author}
  {\bibfnamefont {R.}~\bibnamefont {Sch\"ofbeck}},\ and\ \bibinfo {author}
  {\bibfnamefont {D.}~\bibnamefont {Schwarz}},\ }\bibfield  {title} {\bibinfo
  {title} {{Using the $W$ as a Standard Candle to Reach the Top: Calibrating
  Energy Correlator Based Top Mass Measurements}},\ }\href@noop {} {\
  (\bibinfo {year} {2023}{\natexlab{b}})},\ \Eprint
  {https://arxiv.org/abs/2311.02157} {arXiv:2311.02157 [hep-ph]} \BibitemShut
  {NoStop}%
\bibitem [{\citenamefont {Hayrapetyan}\ \emph {et~al.}(2024)\citenamefont
  {Hayrapetyan} \emph {et~al.}}]{CMS:2024mlf}%
  \BibitemOpen
  \bibfield  {author} {\bibinfo {author} {\bibfnamefont {A.}~\bibnamefont
  {Hayrapetyan}} \emph {et~al.} (\bibinfo {collaboration} {CMS}),\ }\bibfield
  {title} {\bibinfo {title} {{Measurement of energy correlators inside jets and
  determination of the strong coupling $\alpha_\mathrm{S}(m_\mathrm{Z})$}},\
  }\href@noop {} {\  (\bibinfo {year} {2024})},\ \Eprint
  {https://arxiv.org/abs/2402.13864} {arXiv:2402.13864 [hep-ex]} \BibitemShut
  {NoStop}%
\bibitem [{\citenamefont {Abbate}\ \emph {et~al.}(2011)\citenamefont {Abbate},
  \citenamefont {Fickinger}, \citenamefont {Hoang}, \citenamefont {Mateu},\
  and\ \citenamefont {Stewart}}]{Abbate:2010xh}%
  \BibitemOpen
  \bibfield  {author} {\bibinfo {author} {\bibfnamefont {R.}~\bibnamefont
  {Abbate}}, \bibinfo {author} {\bibfnamefont {M.}~\bibnamefont {Fickinger}},
  \bibinfo {author} {\bibfnamefont {A.~H.}\ \bibnamefont {Hoang}}, \bibinfo
  {author} {\bibfnamefont {V.}~\bibnamefont {Mateu}},\ and\ \bibinfo {author}
  {\bibfnamefont {I.~W.}\ \bibnamefont {Stewart}},\ }\bibfield  {title}
  {\bibinfo {title} {{Thrust at $N^{3}LL$ with Power Corrections and a
  Precision Global Fit for $\alpha_{s}(mZ)$}},\ }\href
  {https://doi.org/10.1103/PhysRevD.83.074021} {\bibfield  {journal} {\bibinfo
  {journal} {Phys. Rev. D}\ }\textbf {\bibinfo {volume} {83}},\ \bibinfo
  {pages} {074021} (\bibinfo {year} {2011})},\ \Eprint
  {https://arxiv.org/abs/1006.3080} {arXiv:1006.3080 [hep-ph]} \BibitemShut
  {NoStop}%
\bibitem [{\citenamefont {Mazzilli}(2024)}]{Mazzilli:2024ots}%
  \BibitemOpen
  \bibfield  {author} {\bibinfo {author} {\bibfnamefont {M.}~\bibnamefont
  {Mazzilli}} (\bibinfo {collaboration} {ALICE}),\ }\bibfield  {title}
  {\bibinfo {title} {{Measurements of HF-tagged jet substructure and
  energy-energy correlators with ALICE}},\ }\href
  {https://doi.org/10.22323/1.449.0262} {\bibfield  {journal} {\bibinfo
  {journal} {PoS}\ }\textbf {\bibinfo {volume} {EPS-HEP2023}},\ \bibinfo
  {pages} {262} (\bibinfo {year} {2024})}\BibitemShut {NoStop}%
\bibitem [{\citenamefont {Tamis}(2023)}]{Tamis:2023guc}%
  \BibitemOpen
  \bibfield  {author} {\bibinfo {author} {\bibfnamefont {A.}~\bibnamefont
  {Tamis}},\ }\bibfield  {title} {\bibinfo {title} {{Measurement of Two-Point
  Energy Correlators Within Jets in $pp$ Collisions at $\sqrt{s}$ = 200 GeV at
  STAR}},\ }in\ \href@noop {} {\emph {\bibinfo {booktitle} {{11th International
  Conference on Hard and Electromagnetic Probes of High-Energy Nuclear
  Collisions}: {Hard Probes 2023}}}}\ (\bibinfo {year} {2023})\ \Eprint
  {https://arxiv.org/abs/2309.05761} {arXiv:2309.05761 [hep-ex]} \BibitemShut
  {NoStop}%
\bibitem [{\citenamefont {{ALICE Collaboration (Wenqing
  Fan)}}(2023)}]{Fan2023}%
  \BibitemOpen
  \bibfield  {author} {\bibinfo {author} {\bibnamefont {{ALICE Collaboration
  (Wenqing Fan)}}},\ }\href@noop {} {\bibinfo {title} {First energy-energy
  correlators measurements for inclusive and heavy-flavour tagged jets with
  alice}},\ \bibinfo {howpublished} {Presentation at Quark Matter 2023}
  (\bibinfo {year} {2023}),\ \bibinfo {note} {uRL:
  \url{https://indico.cern.ch/event/1139644/contributions/5541331/attachments/2709459/4704634/QM2023_wide_wenqing_main_Sep5.pdf}}\BibitemShut
  {NoStop}%
\bibitem [{\citenamefont {Chicherin}\ \emph {et~al.}(2021)\citenamefont
  {Chicherin}, \citenamefont {Henn}, \citenamefont {Sokatchev},\ and\
  \citenamefont {Yan}}]{Chicherin:2020azt}%
  \BibitemOpen
  \bibfield  {author} {\bibinfo {author} {\bibfnamefont {D.}~\bibnamefont
  {Chicherin}}, \bibinfo {author} {\bibfnamefont {J.~M.}\ \bibnamefont {Henn}},
  \bibinfo {author} {\bibfnamefont {E.}~\bibnamefont {Sokatchev}},\ and\
  \bibinfo {author} {\bibfnamefont {K.}~\bibnamefont {Yan}},\ }\bibfield
  {title} {\bibinfo {title} {{From correlation functions to event shapes in
  QCD}},\ }\href {https://doi.org/10.1007/JHEP02(2021)053} {\bibfield
  {journal} {\bibinfo  {journal} {JHEP}\ }\textbf {\bibinfo {volume} {02}},\
  \bibinfo {pages} {053}},\ \Eprint {https://arxiv.org/abs/2001.10806}
  {arXiv:2001.10806 [hep-th]} \BibitemShut {NoStop}%
\bibitem [{\citenamefont {Chen}\ \emph {et~al.}(2022)\citenamefont {Chen},
  \citenamefont {Moult}, \citenamefont {Sandor},\ and\ \citenamefont
  {Zhu}}]{Chen:2022jhb}%
  \BibitemOpen
  \bibfield  {author} {\bibinfo {author} {\bibfnamefont {H.}~\bibnamefont
  {Chen}}, \bibinfo {author} {\bibfnamefont {I.}~\bibnamefont {Moult}},
  \bibinfo {author} {\bibfnamefont {J.}~\bibnamefont {Sandor}},\ and\ \bibinfo
  {author} {\bibfnamefont {H.~X.}\ \bibnamefont {Zhu}},\ }\bibfield  {title}
  {\bibinfo {title} {{Celestial Blocks and Transverse Spin in the Three-Point
  Energy Correlator}},\ }\href@noop {} {\  (\bibinfo {year} {2022})},\ \Eprint
  {https://arxiv.org/abs/2202.04085} {arXiv:2202.04085 [hep-ph]} \BibitemShut
  {NoStop}%
\bibitem [{\citenamefont {Chang}\ and\ \citenamefont
  {Simmons-Duffin}(2023)}]{Chang:2022ryc}%
  \BibitemOpen
  \bibfield  {author} {\bibinfo {author} {\bibfnamefont {C.-H.}\ \bibnamefont
  {Chang}}\ and\ \bibinfo {author} {\bibfnamefont {D.}~\bibnamefont
  {Simmons-Duffin}},\ }\bibfield  {title} {\bibinfo {title} {{Three-point
  energy correlators and the celestial block expansion}},\ }\href
  {https://doi.org/10.1007/JHEP02(2023)126} {\bibfield  {journal} {\bibinfo
  {journal} {JHEP}\ }\textbf {\bibinfo {volume} {02}},\ \bibinfo {pages}
  {126}},\ \Eprint {https://arxiv.org/abs/2202.04090} {arXiv:2202.04090
  [hep-th]} \BibitemShut {NoStop}%
\bibitem [{\citenamefont {Chen}\ \emph
  {et~al.}(2023{\natexlab{a}})\citenamefont {Chen}, \citenamefont {Zhou},\ and\
  \citenamefont {Zhu}}]{Chen:2023wah}%
  \BibitemOpen
  \bibfield  {author} {\bibinfo {author} {\bibfnamefont {H.}~\bibnamefont
  {Chen}}, \bibinfo {author} {\bibfnamefont {X.}~\bibnamefont {Zhou}},\ and\
  \bibinfo {author} {\bibfnamefont {H.~X.}\ \bibnamefont {Zhu}},\ }\bibfield
  {title} {\bibinfo {title} {{Power corrections to energy flow correlations
  from large spin perturbation}},\ }\href
  {https://doi.org/10.1007/JHEP10(2023)132} {\bibfield  {journal} {\bibinfo
  {journal} {JHEP}\ }\textbf {\bibinfo {volume} {10}},\ \bibinfo {pages}
  {132}},\ \Eprint {https://arxiv.org/abs/2301.03616} {arXiv:2301.03616
  [hep-ph]} \BibitemShut {NoStop}%
\bibitem [{\citenamefont {Alday}(2017)}]{Alday:2016njk}%
  \BibitemOpen
  \bibfield  {author} {\bibinfo {author} {\bibfnamefont {L.~F.}\ \bibnamefont
  {Alday}},\ }\bibfield  {title} {\bibinfo {title} {{Large Spin Perturbation
  Theory for Conformal Field Theories}},\ }\href
  {https://doi.org/10.1103/PhysRevLett.119.111601} {\bibfield  {journal}
  {\bibinfo  {journal} {Phys. Rev. Lett.}\ }\textbf {\bibinfo {volume} {119}},\
  \bibinfo {pages} {111601} (\bibinfo {year} {2017})},\ \Eprint
  {https://arxiv.org/abs/1611.01500} {arXiv:1611.01500 [hep-th]} \BibitemShut
  {NoStop}%
\bibitem [{\citenamefont {Del~Duca}\ \emph {et~al.}(2016)\citenamefont
  {Del~Duca}, \citenamefont {Duhr}, \citenamefont {Kardos}, \citenamefont
  {Somogyi},\ and\ \citenamefont {Tr\'ocs\'anyi}}]{DelDuca:2016csb}%
  \BibitemOpen
  \bibfield  {author} {\bibinfo {author} {\bibfnamefont {V.}~\bibnamefont
  {Del~Duca}}, \bibinfo {author} {\bibfnamefont {C.}~\bibnamefont {Duhr}},
  \bibinfo {author} {\bibfnamefont {A.}~\bibnamefont {Kardos}}, \bibinfo
  {author} {\bibfnamefont {G.}~\bibnamefont {Somogyi}},\ and\ \bibinfo {author}
  {\bibfnamefont {Z.}~\bibnamefont {Tr\'ocs\'anyi}},\ }\bibfield  {title}
  {\bibinfo {title} {{Three-Jet Production in Electron-Positron Collisions at
  Next-to-Next-to-Leading Order Accuracy}},\ }\href
  {https://doi.org/10.1103/PhysRevLett.117.152004} {\bibfield  {journal}
  {\bibinfo  {journal} {Phys. Rev. Lett.}\ }\textbf {\bibinfo {volume} {117}},\
  \bibinfo {pages} {152004} (\bibinfo {year} {2016})},\ \Eprint
  {https://arxiv.org/abs/1603.08927} {arXiv:1603.08927 [hep-ph]} \BibitemShut
  {NoStop}%
\bibitem [{\citenamefont {Tulip\'ant}\ \emph {et~al.}(2017)\citenamefont
  {Tulip\'ant}, \citenamefont {Kardos},\ and\ \citenamefont
  {Somogyi}}]{Tulipant:2017ybb}%
  \BibitemOpen
  \bibfield  {author} {\bibinfo {author} {\bibfnamefont {Z.}~\bibnamefont
  {Tulip\'ant}}, \bibinfo {author} {\bibfnamefont {A.}~\bibnamefont {Kardos}},\
  and\ \bibinfo {author} {\bibfnamefont {G.}~\bibnamefont {Somogyi}},\
  }\bibfield  {title} {\bibinfo {title} {{Energy\textendash{}energy correlation
  in electron\textendash{}positron annihilation at NNLL + NNLO accuracy}},\
  }\href {https://doi.org/10.1140/epjc/s10052-017-5320-9} {\bibfield  {journal}
  {\bibinfo  {journal} {Eur. Phys. J. C}\ }\textbf {\bibinfo {volume} {77}},\
  \bibinfo {pages} {749} (\bibinfo {year} {2017})},\ \Eprint
  {https://arxiv.org/abs/1708.04093} {arXiv:1708.04093 [hep-ph]} \BibitemShut
  {NoStop}%
\bibitem [{\citenamefont {Dixon}\ \emph {et~al.}(2018)\citenamefont {Dixon},
  \citenamefont {Luo}, \citenamefont {Shtabovenko}, \citenamefont {Yang},\ and\
  \citenamefont {Zhu}}]{Dixon:2018qgp}%
  \BibitemOpen
  \bibfield  {author} {\bibinfo {author} {\bibfnamefont {L.~J.}\ \bibnamefont
  {Dixon}}, \bibinfo {author} {\bibfnamefont {M.-X.}\ \bibnamefont {Luo}},
  \bibinfo {author} {\bibfnamefont {V.}~\bibnamefont {Shtabovenko}}, \bibinfo
  {author} {\bibfnamefont {T.-Z.}\ \bibnamefont {Yang}},\ and\ \bibinfo
  {author} {\bibfnamefont {H.~X.}\ \bibnamefont {Zhu}},\ }\bibfield  {title}
  {\bibinfo {title} {{Analytical Computation of Energy-Energy Correlation at
  Next-to-Leading Order in QCD}},\ }\href
  {https://doi.org/10.1103/PhysRevLett.120.102001} {\bibfield  {journal}
  {\bibinfo  {journal} {Phys. Rev. Lett.}\ }\textbf {\bibinfo {volume} {120}},\
  \bibinfo {pages} {102001} (\bibinfo {year} {2018})},\ \Eprint
  {https://arxiv.org/abs/1801.03219} {arXiv:1801.03219 [hep-ph]} \BibitemShut
  {NoStop}%
\bibitem [{\citenamefont {Korchemsky}(2020)}]{Korchemsky:2019nzm}%
  \BibitemOpen
  \bibfield  {author} {\bibinfo {author} {\bibfnamefont {G.~P.}\ \bibnamefont
  {Korchemsky}},\ }\bibfield  {title} {\bibinfo {title} {{Energy correlations
  in the end-point region}},\ }\href {https://doi.org/10.1007/JHEP01(2020)008}
  {\bibfield  {journal} {\bibinfo  {journal} {JHEP}\ }\textbf {\bibinfo
  {volume} {01}},\ \bibinfo {pages} {008}},\ \Eprint
  {https://arxiv.org/abs/1905.01444} {arXiv:1905.01444 [hep-th]} \BibitemShut
  {NoStop}%
\bibitem [{\citenamefont {Chen}\ \emph {et~al.}(2021)\citenamefont {Chen},
  \citenamefont {Yang}, \citenamefont {Zhu},\ and\ \citenamefont
  {Zhu}}]{Chen:2020uvt}%
  \BibitemOpen
  \bibfield  {author} {\bibinfo {author} {\bibfnamefont {H.}~\bibnamefont
  {Chen}}, \bibinfo {author} {\bibfnamefont {T.-Z.}\ \bibnamefont {Yang}},
  \bibinfo {author} {\bibfnamefont {H.~X.}\ \bibnamefont {Zhu}},\ and\ \bibinfo
  {author} {\bibfnamefont {Y.~J.}\ \bibnamefont {Zhu}},\ }\bibfield  {title}
  {\bibinfo {title} {{Analytic Continuation and Reciprocity Relation for
  Collinear Splitting in QCD}},\ }\href
  {https://doi.org/10.1088/1674-1137/abde2d} {\bibfield  {journal} {\bibinfo
  {journal} {Chin. Phys. C}\ }\textbf {\bibinfo {volume} {45}},\ \bibinfo
  {pages} {043101} (\bibinfo {year} {2021})},\ \Eprint
  {https://arxiv.org/abs/2006.10534} {arXiv:2006.10534 [hep-ph]} \BibitemShut
  {NoStop}%
\bibitem [{\citenamefont {Kodaira}\ and\ \citenamefont
  {Trentadue}(1982)}]{Kodaira:1981nh}%
  \BibitemOpen
  \bibfield  {author} {\bibinfo {author} {\bibfnamefont {J.}~\bibnamefont
  {Kodaira}}\ and\ \bibinfo {author} {\bibfnamefont {L.}~\bibnamefont
  {Trentadue}},\ }\bibfield  {title} {\bibinfo {title} {{Summing Soft Emission
  in QCD}},\ }\href {https://doi.org/10.1016/0370-2693(82)90907-8} {\bibfield
  {journal} {\bibinfo  {journal} {Phys. Lett. B}\ }\textbf {\bibinfo {volume}
  {112}},\ \bibinfo {pages} {66} (\bibinfo {year} {1982})}\BibitemShut
  {NoStop}%
\bibitem [{\citenamefont {Kodaira}\ and\ \citenamefont
  {Trentadue}(1983)}]{Kodaira:1982az}%
  \BibitemOpen
  \bibfield  {author} {\bibinfo {author} {\bibfnamefont {J.}~\bibnamefont
  {Kodaira}}\ and\ \bibinfo {author} {\bibfnamefont {L.}~\bibnamefont
  {Trentadue}},\ }\bibfield  {title} {\bibinfo {title} {{Single Logarithm
  Effects in electron-Positron Annihilation}},\ }\href
  {https://doi.org/10.1016/0370-2693(83)91213-3} {\bibfield  {journal}
  {\bibinfo  {journal} {Phys. Lett. B}\ }\textbf {\bibinfo {volume} {123}},\
  \bibinfo {pages} {335} (\bibinfo {year} {1983})}\BibitemShut {NoStop}%
\bibitem [{\citenamefont {de~Florian}\ and\ \citenamefont
  {Grazzini}(2005)}]{deFlorian:2004mp}%
  \BibitemOpen
  \bibfield  {author} {\bibinfo {author} {\bibfnamefont {D.}~\bibnamefont
  {de~Florian}}\ and\ \bibinfo {author} {\bibfnamefont {M.}~\bibnamefont
  {Grazzini}},\ }\bibfield  {title} {\bibinfo {title} {{The Back-to-back region
  in e+ e- energy-energy correlation}},\ }\href
  {https://doi.org/10.1016/j.nuclphysb.2004.10.051} {\bibfield  {journal}
  {\bibinfo  {journal} {Nucl. Phys. B}\ }\textbf {\bibinfo {volume} {704}},\
  \bibinfo {pages} {387} (\bibinfo {year} {2005})},\ \Eprint
  {https://arxiv.org/abs/hep-ph/0407241} {arXiv:hep-ph/0407241} \BibitemShut
  {NoStop}%
\bibitem [{\citenamefont {Moult}\ and\ \citenamefont
  {Zhu}(2018)}]{Moult:2018jzp}%
  \BibitemOpen
  \bibfield  {author} {\bibinfo {author} {\bibfnamefont {I.}~\bibnamefont
  {Moult}}\ and\ \bibinfo {author} {\bibfnamefont {H.~X.}\ \bibnamefont
  {Zhu}},\ }\bibfield  {title} {\bibinfo {title} {{Simplicity from Recoil: The
  Three-Loop Soft Function and Factorization for the Energy-Energy
  Correlation}},\ }\href {https://doi.org/10.1007/JHEP08(2018)160} {\bibfield
  {journal} {\bibinfo  {journal} {JHEP}\ }\textbf {\bibinfo {volume} {08}},\
  \bibinfo {pages} {160}},\ \Eprint {https://arxiv.org/abs/1801.02627}
  {arXiv:1801.02627 [hep-ph]} \BibitemShut {NoStop}%
\bibitem [{\citenamefont {Ebert}\ \emph {et~al.}(2021)\citenamefont {Ebert},
  \citenamefont {Mistlberger},\ and\ \citenamefont {Vita}}]{Ebert:2020sfi}%
  \BibitemOpen
  \bibfield  {author} {\bibinfo {author} {\bibfnamefont {M.~A.}\ \bibnamefont
  {Ebert}}, \bibinfo {author} {\bibfnamefont {B.}~\bibnamefont {Mistlberger}},\
  and\ \bibinfo {author} {\bibfnamefont {G.}~\bibnamefont {Vita}},\ }\bibfield
  {title} {\bibinfo {title} {{The Energy-Energy Correlation in the back-to-back
  limit at N$^{3}$LO and N$^{3}$LL'}},\ }\href
  {https://doi.org/10.1007/JHEP08(2021)022} {\bibfield  {journal} {\bibinfo
  {journal} {JHEP}\ }\textbf {\bibinfo {volume} {08}},\ \bibinfo {pages}
  {022}},\ \Eprint {https://arxiv.org/abs/2012.07859} {arXiv:2012.07859
  [hep-ph]} \BibitemShut {NoStop}%
\bibitem [{\citenamefont {Duhr}\ \emph {et~al.}(2022)\citenamefont {Duhr},
  \citenamefont {Mistlberger},\ and\ \citenamefont {Vita}}]{Duhr:2022yyp}%
  \BibitemOpen
  \bibfield  {author} {\bibinfo {author} {\bibfnamefont {C.}~\bibnamefont
  {Duhr}}, \bibinfo {author} {\bibfnamefont {B.}~\bibnamefont {Mistlberger}},\
  and\ \bibinfo {author} {\bibfnamefont {G.}~\bibnamefont {Vita}},\ }\bibfield
  {title} {\bibinfo {title} {{Four-Loop Rapidity Anomalous Dimension and Event
  Shapes to Fourth Logarithmic Order}},\ }\href
  {https://doi.org/10.1103/PhysRevLett.129.162001} {\bibfield  {journal}
  {\bibinfo  {journal} {Phys. Rev. Lett.}\ }\textbf {\bibinfo {volume} {129}},\
  \bibinfo {pages} {162001} (\bibinfo {year} {2022})},\ \Eprint
  {https://arxiv.org/abs/2205.02242} {arXiv:2205.02242 [hep-ph]} \BibitemShut
  {NoStop}%
\bibitem [{\citenamefont {Belitsky}\ \emph
  {et~al.}(2014{\natexlab{c}})\citenamefont {Belitsky}, \citenamefont
  {Hohenegger}, \citenamefont {Korchemsky}, \citenamefont {Sokatchev},\ and\
  \citenamefont {Zhiboedov}}]{Belitsky:2013ofa}%
  \BibitemOpen
  \bibfield  {author} {\bibinfo {author} {\bibfnamefont {A.}~\bibnamefont
  {Belitsky}}, \bibinfo {author} {\bibfnamefont {S.}~\bibnamefont
  {Hohenegger}}, \bibinfo {author} {\bibfnamefont {G.}~\bibnamefont
  {Korchemsky}}, \bibinfo {author} {\bibfnamefont {E.}~\bibnamefont
  {Sokatchev}},\ and\ \bibinfo {author} {\bibfnamefont {A.}~\bibnamefont
  {Zhiboedov}},\ }\bibfield  {title} {\bibinfo {title} {{Energy-Energy
  Correlations in N=4 Supersymmetric Yang-Mills Theory}},\ }\href
  {https://doi.org/10.1103/PhysRevLett.112.071601} {\bibfield  {journal}
  {\bibinfo  {journal} {Phys. Rev. Lett.}\ }\textbf {\bibinfo {volume} {112}},\
  \bibinfo {pages} {071601} (\bibinfo {year} {2014}{\natexlab{c}})},\ \Eprint
  {https://arxiv.org/abs/1311.6800} {arXiv:1311.6800 [hep-th]} \BibitemShut
  {NoStop}%
\bibitem [{\citenamefont {Henn}\ \emph {et~al.}(2019)\citenamefont {Henn},
  \citenamefont {Sokatchev}, \citenamefont {Yan},\ and\ \citenamefont
  {Zhiboedov}}]{Henn:2019gkr}%
  \BibitemOpen
  \bibfield  {author} {\bibinfo {author} {\bibfnamefont {J.~M.}\ \bibnamefont
  {Henn}}, \bibinfo {author} {\bibfnamefont {E.}~\bibnamefont {Sokatchev}},
  \bibinfo {author} {\bibfnamefont {K.}~\bibnamefont {Yan}},\ and\ \bibinfo
  {author} {\bibfnamefont {A.}~\bibnamefont {Zhiboedov}},\ }\bibfield  {title}
  {\bibinfo {title} {{Energy-energy correlation in $N$=4 super Yang-Mills
  theory at next-to-next-to-leading order}},\ }\href
  {https://doi.org/10.1103/PhysRevD.100.036010} {\bibfield  {journal} {\bibinfo
   {journal} {Phys. Rev. D}\ }\textbf {\bibinfo {volume} {100}},\ \bibinfo
  {pages} {036010} (\bibinfo {year} {2019})},\ \Eprint
  {https://arxiv.org/abs/1903.05314} {arXiv:1903.05314 [hep-th]} \BibitemShut
  {NoStop}%
\bibitem [{\citenamefont {Moult}\ \emph {et~al.}(2020)\citenamefont {Moult},
  \citenamefont {Vita},\ and\ \citenamefont {Yan}}]{Moult:2019vou}%
  \BibitemOpen
  \bibfield  {author} {\bibinfo {author} {\bibfnamefont {I.}~\bibnamefont
  {Moult}}, \bibinfo {author} {\bibfnamefont {G.}~\bibnamefont {Vita}},\ and\
  \bibinfo {author} {\bibfnamefont {K.}~\bibnamefont {Yan}},\ }\bibfield
  {title} {\bibinfo {title} {{Subleading power resummation of rapidity
  logarithms: the energy-energy correlator in $ \mathcal{N} $ = 4 SYM}},\
  }\href {https://doi.org/10.1007/JHEP07(2020)005} {\bibfield  {journal}
  {\bibinfo  {journal} {JHEP}\ }\textbf {\bibinfo {volume} {07}},\ \bibinfo
  {pages} {005}},\ \Eprint {https://arxiv.org/abs/1912.02188} {arXiv:1912.02188
  [hep-ph]} \BibitemShut {NoStop}%
\bibitem [{\citenamefont {Kologlu}\ \emph {et~al.}(2021)\citenamefont
  {Kologlu}, \citenamefont {Kravchuk}, \citenamefont {Simmons-Duffin},\ and\
  \citenamefont {Zhiboedov}}]{Kologlu:2019mfz}%
  \BibitemOpen
  \bibfield  {author} {\bibinfo {author} {\bibfnamefont {M.}~\bibnamefont
  {Kologlu}}, \bibinfo {author} {\bibfnamefont {P.}~\bibnamefont {Kravchuk}},
  \bibinfo {author} {\bibfnamefont {D.}~\bibnamefont {Simmons-Duffin}},\ and\
  \bibinfo {author} {\bibfnamefont {A.}~\bibnamefont {Zhiboedov}},\ }\bibfield
  {title} {\bibinfo {title} {{The light-ray OPE and conformal colliders}},\
  }\href {https://doi.org/10.1007/JHEP01(2021)128} {\bibfield  {journal}
  {\bibinfo  {journal} {JHEP}\ }\textbf {\bibinfo {volume} {01}},\ \bibinfo
  {pages} {128}},\ \Eprint {https://arxiv.org/abs/1905.01311} {arXiv:1905.01311
  [hep-th]} \BibitemShut {NoStop}%
\bibitem [{\citenamefont {Lee}\ and\ \citenamefont
  {Sterman}(2007)}]{Lee:2006nr}%
  \BibitemOpen
  \bibfield  {author} {\bibinfo {author} {\bibfnamefont {C.}~\bibnamefont
  {Lee}}\ and\ \bibinfo {author} {\bibfnamefont {G.~F.}\ \bibnamefont
  {Sterman}},\ }\bibfield  {title} {\bibinfo {title} {{Momentum Flow
  Correlations from Event Shapes: Factorized Soft Gluons and Soft-Collinear
  Effective Theory}},\ }\href {https://doi.org/10.1103/PhysRevD.75.014022}
  {\bibfield  {journal} {\bibinfo  {journal} {Phys. Rev. D}\ }\textbf {\bibinfo
  {volume} {75}},\ \bibinfo {pages} {014022} (\bibinfo {year} {2007})},\
  \Eprint {https://arxiv.org/abs/hep-ph/0611061} {arXiv:hep-ph/0611061}
  \BibitemShut {NoStop}%
\bibitem [{\citenamefont {Stewart}\ \emph {et~al.}(2015)\citenamefont
  {Stewart}, \citenamefont {Tackmann},\ and\ \citenamefont
  {Waalewijn}}]{Stewart:2014nna}%
  \BibitemOpen
  \bibfield  {author} {\bibinfo {author} {\bibfnamefont {I.~W.}\ \bibnamefont
  {Stewart}}, \bibinfo {author} {\bibfnamefont {F.~J.}\ \bibnamefont
  {Tackmann}},\ and\ \bibinfo {author} {\bibfnamefont {W.~J.}\ \bibnamefont
  {Waalewijn}},\ }\bibfield  {title} {\bibinfo {title} {{Dissecting Soft
  Radiation with Factorization}},\ }\href
  {https://doi.org/10.1103/PhysRevLett.114.092001} {\bibfield  {journal}
  {\bibinfo  {journal} {Phys. Rev. Lett.}\ }\textbf {\bibinfo {volume} {114}},\
  \bibinfo {pages} {092001} (\bibinfo {year} {2015})},\ \Eprint
  {https://arxiv.org/abs/1405.6722} {arXiv:1405.6722 [hep-ph]} \BibitemShut
  {NoStop}%
\bibitem [{\citenamefont {Ferdinand}\ \emph {et~al.}(2023)\citenamefont
  {Ferdinand}, \citenamefont {Lee},\ and\ \citenamefont
  {Pathak}}]{Ferdinand:2023vaf}%
  \BibitemOpen
  \bibfield  {author} {\bibinfo {author} {\bibfnamefont {A.}~\bibnamefont
  {Ferdinand}}, \bibinfo {author} {\bibfnamefont {K.}~\bibnamefont {Lee}},\
  and\ \bibinfo {author} {\bibfnamefont {A.}~\bibnamefont {Pathak}},\
  }\bibfield  {title} {\bibinfo {title} {{Field-theoretic analysis of
  hadronization using soft drop jet mass}},\ }\href
  {https://doi.org/10.1103/PhysRevD.108.L111501} {\bibfield  {journal}
  {\bibinfo  {journal} {Phys. Rev. D}\ }\textbf {\bibinfo {volume} {108}},\
  \bibinfo {pages} {L111501} (\bibinfo {year} {2023})},\ \Eprint
  {https://arxiv.org/abs/2301.03605} {arXiv:2301.03605 [hep-ph]} \BibitemShut
  {NoStop}%
\bibitem [{\citenamefont {Dokshitzer}\ and\ \citenamefont
  {Webber}(1995)}]{Dokshitzer:1995zt}%
  \BibitemOpen
  \bibfield  {author} {\bibinfo {author} {\bibfnamefont {Y.~L.}\ \bibnamefont
  {Dokshitzer}}\ and\ \bibinfo {author} {\bibfnamefont {B.~R.}\ \bibnamefont
  {Webber}},\ }\bibfield  {title} {\bibinfo {title} {{Calculation of power
  corrections to hadronic event shapes}},\ }\href
  {https://doi.org/10.1016/0370-2693(95)00548-Y} {\bibfield  {journal}
  {\bibinfo  {journal} {Phys. Lett.}\ }\textbf {\bibinfo {volume} {B352}},\
  \bibinfo {pages} {451} (\bibinfo {year} {1995})},\ \Eprint
  {https://arxiv.org/abs/hep-ph/9504219} {arXiv:hep-ph/9504219} \BibitemShut
  {NoStop}%
\bibitem [{\citenamefont {Gardi}(2000)}]{Gardi:2000yh}%
  \BibitemOpen
  \bibfield  {author} {\bibinfo {author} {\bibfnamefont {E.}~\bibnamefont
  {Gardi}},\ }\bibfield  {title} {\bibinfo {title} {{Perturbative and
  nonperturbative aspects of moments of the thrust distribution in e+ e-
  annihilation}},\ }\href {https://doi.org/10.1088/1126-6708/2000/04/030}
  {\bibfield  {journal} {\bibinfo  {journal} {JHEP}\ }\textbf {\bibinfo
  {volume} {04}},\ \bibinfo {pages} {030}},\ \Eprint
  {https://arxiv.org/abs/hep-ph/0003179} {arXiv:hep-ph/0003179 [hep-ph]}
  \BibitemShut {NoStop}%
\bibitem [{\citenamefont {Mateu}\ \emph {et~al.}(2013)\citenamefont {Mateu},
  \citenamefont {Stewart},\ and\ \citenamefont {Thaler}}]{Mateu:2012nk}%
  \BibitemOpen
  \bibfield  {author} {\bibinfo {author} {\bibfnamefont {V.}~\bibnamefont
  {Mateu}}, \bibinfo {author} {\bibfnamefont {I.~W.}\ \bibnamefont {Stewart}},\
  and\ \bibinfo {author} {\bibfnamefont {J.}~\bibnamefont {Thaler}},\
  }\bibfield  {title} {\bibinfo {title} {{Power Corrections to Event Shapes
  with Mass-Dependent Operators}},\ }\href
  {https://doi.org/10.1103/PhysRevD.87.014025} {\bibfield  {journal} {\bibinfo
  {journal} {Phys. Rev. D}\ }\textbf {\bibinfo {volume} {87}},\ \bibinfo
  {pages} {014025} (\bibinfo {year} {2013})},\ \Eprint
  {https://arxiv.org/abs/1209.3781} {arXiv:1209.3781 [hep-ph]} \BibitemShut
  {NoStop}%
\bibitem [{\citenamefont {Schindler}\ \emph {et~al.}(2023)\citenamefont
  {Schindler}, \citenamefont {Stewart},\ and\ \citenamefont
  {Sun}}]{Schindler:2023cww}%
  \BibitemOpen
  \bibfield  {author} {\bibinfo {author} {\bibfnamefont {S.~T.}\ \bibnamefont
  {Schindler}}, \bibinfo {author} {\bibfnamefont {I.~W.}\ \bibnamefont
  {Stewart}},\ and\ \bibinfo {author} {\bibfnamefont {Z.}~\bibnamefont {Sun}},\
  }\bibfield  {title} {\bibinfo {title} {{Renormalons in the energy-energy
  correlator}},\ }\href {https://doi.org/10.1007/JHEP10(2023)187} {\bibfield
  {journal} {\bibinfo  {journal} {JHEP}\ }\textbf {\bibinfo {volume} {10}},\
  \bibinfo {pages} {187}},\ \Eprint {https://arxiv.org/abs/2305.19311}
  {arXiv:2305.19311 [hep-ph]} \BibitemShut {NoStop}%
\bibitem [{\citenamefont {Akrawy}\ \emph {et~al.}(1990)\citenamefont {Akrawy}
  \emph {et~al.}}]{OPAL:1990reb}%
  \BibitemOpen
  \bibfield  {author} {\bibinfo {author} {\bibfnamefont {M.~Z.}\ \bibnamefont
  {Akrawy}} \emph {et~al.} (\bibinfo {collaboration} {OPAL}),\ }\bibfield
  {title} {\bibinfo {title} {{A Measurement of energy correlations and a
  determination of alpha-s (M2 (Z0)) in e+ e- annihilations at s**(1/2) =
  91-GeV}},\ }\href {https://doi.org/10.1016/0370-2693(90)91098-V} {\bibfield
  {journal} {\bibinfo  {journal} {Phys. Lett. B}\ }\textbf {\bibinfo {volume}
  {252}},\ \bibinfo {pages} {159} (\bibinfo {year} {1990})}\BibitemShut
  {NoStop}%
\bibitem [{\citenamefont {Decamp}\ \emph {et~al.}(1991)\citenamefont {Decamp}
  \emph {et~al.}}]{ALEPH:1990vew}%
  \BibitemOpen
  \bibfield  {author} {\bibinfo {author} {\bibfnamefont {D.}~\bibnamefont
  {Decamp}} \emph {et~al.} (\bibinfo {collaboration} {ALEPH}),\ }\bibfield
  {title} {\bibinfo {title} {{Measurement of alpha-s from the structure of
  particle clusters produced in hadronic Z decays}},\ }\href
  {https://doi.org/10.1016/0370-2693(91)91926-M} {\bibfield  {journal}
  {\bibinfo  {journal} {Phys. Lett. B}\ }\textbf {\bibinfo {volume} {257}},\
  \bibinfo {pages} {479} (\bibinfo {year} {1991})}\BibitemShut {NoStop}%
\bibitem [{\citenamefont {Adeva}\ \emph {et~al.}(1991)\citenamefont {Adeva}
  \emph {et~al.}}]{L3:1991qlf}%
  \BibitemOpen
  \bibfield  {author} {\bibinfo {author} {\bibfnamefont {B.}~\bibnamefont
  {Adeva}} \emph {et~al.} (\bibinfo {collaboration} {L3}),\ }\bibfield  {title}
  {\bibinfo {title} {{Determination of alpha-s from energy-energy correlations
  measured on the Z0 resonance.}},\ }\href
  {https://doi.org/10.1016/0370-2693(91)91925-L} {\bibfield  {journal}
  {\bibinfo  {journal} {Phys. Lett. B}\ }\textbf {\bibinfo {volume} {257}},\
  \bibinfo {pages} {469} (\bibinfo {year} {1991})}\BibitemShut {NoStop}%
\bibitem [{\citenamefont {Acton}\ \emph {et~al.}(1993)\citenamefont {Acton}
  \emph {et~al.}}]{OPAL:1993pnw}%
  \BibitemOpen
  \bibfield  {author} {\bibinfo {author} {\bibfnamefont {P.~D.}\ \bibnamefont
  {Acton}} \emph {et~al.} (\bibinfo {collaboration} {OPAL}),\ }\bibfield
  {title} {\bibinfo {title} {{A Determination of alpha-s (M (Z0)) at LEP using
  resummed QCD calculations}},\ }\href {https://doi.org/10.1007/BF01555834}
  {\bibfield  {journal} {\bibinfo  {journal} {Z. Phys. C}\ }\textbf {\bibinfo
  {volume} {59}},\ \bibinfo {pages} {1} (\bibinfo {year} {1993})}\BibitemShut
  {NoStop}%
\bibitem [{\citenamefont {Beneke}(1999)}]{Beneke:1998ui}%
  \BibitemOpen
  \bibfield  {author} {\bibinfo {author} {\bibfnamefont {M.}~\bibnamefont
  {Beneke}},\ }\bibfield  {title} {\bibinfo {title} {{Renormalons}},\ }\href
  {https://doi.org/10.1016/S0370-1573(98)00130-6} {\bibfield  {journal}
  {\bibinfo  {journal} {Phys. Rept.}\ }\textbf {\bibinfo {volume} {317}},\
  \bibinfo {pages} {1} (\bibinfo {year} {1999})},\ \Eprint
  {https://arxiv.org/abs/hep-ph/9807443} {arXiv:hep-ph/9807443} \BibitemShut
  {NoStop}%
\bibitem [{\citenamefont {Argyres}\ and\ \citenamefont
  {Unsal}(2012)}]{Argyres:2012ka}%
  \BibitemOpen
  \bibfield  {author} {\bibinfo {author} {\bibfnamefont {P.~C.}\ \bibnamefont
  {Argyres}}\ and\ \bibinfo {author} {\bibfnamefont {M.}~\bibnamefont
  {Unsal}},\ }\bibfield  {title} {\bibinfo {title} {{The semi-classical
  expansion and resurgence in gauge theories: new perturbative, instanton,
  bion, and renormalon effects}},\ }\href
  {https://doi.org/10.1007/JHEP08(2012)063} {\bibfield  {journal} {\bibinfo
  {journal} {JHEP}\ }\textbf {\bibinfo {volume} {08}},\ \bibinfo {pages}
  {063}},\ \Eprint {https://arxiv.org/abs/1206.1890} {arXiv:1206.1890 [hep-th]}
  \BibitemShut {NoStop}%
\bibitem [{\citenamefont {Dunne}\ and\ \citenamefont
  {\"Unsal}(2014)}]{Dunne:2013ada}%
  \BibitemOpen
  \bibfield  {author} {\bibinfo {author} {\bibfnamefont {G.~V.}\ \bibnamefont
  {Dunne}}\ and\ \bibinfo {author} {\bibfnamefont {M.}~\bibnamefont
  {\"Unsal}},\ }\bibfield  {title} {\bibinfo {title} {{Generating
  nonperturbative physics from perturbation theory}},\ }\href
  {https://doi.org/10.1103/PhysRevD.89.041701} {\bibfield  {journal} {\bibinfo
  {journal} {Phys. Rev. D}\ }\textbf {\bibinfo {volume} {89}},\ \bibinfo
  {pages} {041701} (\bibinfo {year} {2014})},\ \Eprint
  {https://arxiv.org/abs/1306.4405} {arXiv:1306.4405 [hep-th]} \BibitemShut
  {NoStop}%
\bibitem [{\citenamefont {Gross}\ and\ \citenamefont
  {Neveu}(1974)}]{Gross:1974jv}%
  \BibitemOpen
  \bibfield  {author} {\bibinfo {author} {\bibfnamefont {D.~J.}\ \bibnamefont
  {Gross}}\ and\ \bibinfo {author} {\bibfnamefont {A.}~\bibnamefont {Neveu}},\
  }\bibfield  {title} {\bibinfo {title} {{Dynamical Symmetry Breaking in
  Asymptotically Free Field Theories}},\ }\href
  {https://doi.org/10.1103/PhysRevD.10.3235} {\bibfield  {journal} {\bibinfo
  {journal} {Phys. Rev. D}\ }\textbf {\bibinfo {volume} {10}},\ \bibinfo
  {pages} {3235} (\bibinfo {year} {1974})}\BibitemShut {NoStop}%
\bibitem [{\citenamefont {Lautrup}(1977)}]{Lautrup:1977hs}%
  \BibitemOpen
  \bibfield  {author} {\bibinfo {author} {\bibfnamefont {B.~E.}\ \bibnamefont
  {Lautrup}},\ }\bibfield  {title} {\bibinfo {title} {{On High Order Estimates
  in QED}},\ }\href {https://doi.org/10.1016/0370-2693(77)90145-9} {\bibfield
  {journal} {\bibinfo  {journal} {Phys. Lett. B}\ }\textbf {\bibinfo {volume}
  {69}},\ \bibinfo {pages} {109} (\bibinfo {year} {1977})}\BibitemShut
  {NoStop}%
\bibitem [{\citenamefont {'t~Hooft}(1979)}]{tHooft:1977xjm}%
  \BibitemOpen
  \bibfield  {author} {\bibinfo {author} {\bibfnamefont {G.}~\bibnamefont
  {'t~Hooft}},\ }\bibfield  {title} {\bibinfo {title} {{Can We Make Sense Out
  of Quantum Chromodynamics?}},\ }\href@noop {} {\bibfield  {journal} {\bibinfo
   {journal} {Subnucl. Ser.}\ }\textbf {\bibinfo {volume} {15}},\ \bibinfo
  {pages} {943} (\bibinfo {year} {1979})}\BibitemShut {NoStop}%
\bibitem [{\citenamefont {Hoang}\ and\ \citenamefont
  {Stewart}(2008)}]{Hoang:2007vb}%
  \BibitemOpen
  \bibfield  {author} {\bibinfo {author} {\bibfnamefont {A.~H.}\ \bibnamefont
  {Hoang}}\ and\ \bibinfo {author} {\bibfnamefont {I.~W.}\ \bibnamefont
  {Stewart}},\ }\bibfield  {title} {\bibinfo {title} {{Designing Gapped Soft
  Functions for Jet Production}},\ }\href
  {https://doi.org/10.1016/j.physletb.2008.01.040} {\bibfield  {journal}
  {\bibinfo  {journal} {Phys. Lett.}\ }\textbf {\bibinfo {volume} {B660}},\
  \bibinfo {pages} {483} (\bibinfo {year} {2008})},\ \Eprint
  {https://arxiv.org/abs/0709.3519} {arXiv:0709.3519 [hep-ph]} \BibitemShut
  {NoStop}%
\bibitem [{\citenamefont {Hoang}\ and\ \citenamefont
  {Kluth}(2008)}]{Hoang:2008fs}%
  \BibitemOpen
  \bibfield  {author} {\bibinfo {author} {\bibfnamefont {A.~H.}\ \bibnamefont
  {Hoang}}\ and\ \bibinfo {author} {\bibfnamefont {S.}~\bibnamefont {Kluth}},\
  }\bibfield  {title} {\bibinfo {title} {{Hemisphere Soft Function at
  $O(\alpha_s^2)$ for Dijet Production in $e^+e^-$ Annihilation}},\ }\href@noop
  {} {\  (\bibinfo {year} {2008})},\ \Eprint {https://arxiv.org/abs/0806.3852}
  {arXiv:0806.3852 [hep-ph]} \BibitemShut {NoStop}%
\bibitem [{\citenamefont {Hoang}\ \emph {et~al.}(2010)\citenamefont {Hoang},
  \citenamefont {Jain}, \citenamefont {Scimemi},\ and\ \citenamefont
  {Stewart}}]{Hoang:2009yr}%
  \BibitemOpen
  \bibfield  {author} {\bibinfo {author} {\bibfnamefont {A.~H.}\ \bibnamefont
  {Hoang}}, \bibinfo {author} {\bibfnamefont {A.}~\bibnamefont {Jain}},
  \bibinfo {author} {\bibfnamefont {I.}~\bibnamefont {Scimemi}},\ and\ \bibinfo
  {author} {\bibfnamefont {I.~W.}\ \bibnamefont {Stewart}},\ }\bibfield
  {title} {\bibinfo {title} {{R-evolution: Improving perturbative QCD}},\
  }\href {https://doi.org/10.1103/PhysRevD.82.011501} {\bibfield  {journal}
  {\bibinfo  {journal} {Phys. Rev. D}\ }\textbf {\bibinfo {volume} {82}},\
  \bibinfo {pages} {011501} (\bibinfo {year} {2010})},\ \Eprint
  {https://arxiv.org/abs/0908.3189} {arXiv:0908.3189 [hep-ph]} \BibitemShut
  {NoStop}%
\bibitem [{\citenamefont {Bachu}\ \emph {et~al.}(2021)\citenamefont {Bachu},
  \citenamefont {Hoang}, \citenamefont {Mateu}, \citenamefont {Pathak},\ and\
  \citenamefont {Stewart}}]{Bachu:2020nqn}%
  \BibitemOpen
  \bibfield  {author} {\bibinfo {author} {\bibfnamefont {B.}~\bibnamefont
  {Bachu}}, \bibinfo {author} {\bibfnamefont {A.~H.}\ \bibnamefont {Hoang}},
  \bibinfo {author} {\bibfnamefont {V.}~\bibnamefont {Mateu}}, \bibinfo
  {author} {\bibfnamefont {A.}~\bibnamefont {Pathak}},\ and\ \bibinfo {author}
  {\bibfnamefont {I.~W.}\ \bibnamefont {Stewart}},\ }\bibfield  {title}
  {\bibinfo {title} {{Boosted top quarks in the peak region with NL3L
  resummation}},\ }\href {https://doi.org/10.1103/PhysRevD.104.014026}
  {\bibfield  {journal} {\bibinfo  {journal} {Phys. Rev. D}\ }\textbf {\bibinfo
  {volume} {104}},\ \bibinfo {pages} {014026} (\bibinfo {year} {2021})},\
  \Eprint {https://arxiv.org/abs/2012.12304} {arXiv:2012.12304 [hep-ph]}
  \BibitemShut {NoStop}%
\bibitem [{\citenamefont {Hoang}\ \emph {et~al.}(2015)\citenamefont {Hoang},
  \citenamefont {Kolodrubetz}, \citenamefont {Mateu},\ and\ \citenamefont
  {Stewart}}]{Hoang:2014wka}%
  \BibitemOpen
  \bibfield  {author} {\bibinfo {author} {\bibfnamefont {A.~H.}\ \bibnamefont
  {Hoang}}, \bibinfo {author} {\bibfnamefont {D.~W.}\ \bibnamefont
  {Kolodrubetz}}, \bibinfo {author} {\bibfnamefont {V.}~\bibnamefont {Mateu}},\
  and\ \bibinfo {author} {\bibfnamefont {I.~W.}\ \bibnamefont {Stewart}},\
  }\bibfield  {title} {\bibinfo {title} {{$C$-parameter distribution at
  N$^3$LL' including power corrections}},\ }\href
  {https://doi.org/10.1103/PhysRevD.91.094017} {\bibfield  {journal} {\bibinfo
  {journal} {Phys. Rev. D}\ }\textbf {\bibinfo {volume} {91}},\ \bibinfo
  {pages} {094017} (\bibinfo {year} {2015})},\ \Eprint
  {https://arxiv.org/abs/1411.6633} {arXiv:1411.6633 [hep-ph]} \BibitemShut
  {NoStop}%
\bibitem [{\citenamefont {Dehnadi}\ \emph {et~al.}(2023)\citenamefont
  {Dehnadi}, \citenamefont {Hoang}, \citenamefont {Jin},\ and\ \citenamefont
  {Mateu}}]{Dehnadi:2023msm}%
  \BibitemOpen
  \bibfield  {author} {\bibinfo {author} {\bibfnamefont {B.}~\bibnamefont
  {Dehnadi}}, \bibinfo {author} {\bibfnamefont {A.~H.}\ \bibnamefont {Hoang}},
  \bibinfo {author} {\bibfnamefont {O.~L.}\ \bibnamefont {Jin}},\ and\ \bibinfo
  {author} {\bibfnamefont {V.}~\bibnamefont {Mateu}},\ }\bibfield  {title}
  {\bibinfo {title} {{Top quark mass calibration for Monte Carlo event
  generators \textemdash{} an update}},\ }\href
  {https://doi.org/10.1007/JHEP12(2023)065} {\bibfield  {journal} {\bibinfo
  {journal} {JHEP}\ }\textbf {\bibinfo {volume} {12}},\ \bibinfo {pages}
  {065}},\ \Eprint {https://arxiv.org/abs/2309.00547} {arXiv:2309.00547
  [hep-ph]} \BibitemShut {NoStop}%
\bibitem [{\citenamefont {Lee}\ \emph {et~al.}()\citenamefont {Lee},
  \citenamefont {Pathak}, \citenamefont {Stewart},\ and\ \citenamefont
  {Sun}}]{LPSS-Long}%
  \BibitemOpen
  \bibfield  {author} {\bibinfo {author} {\bibfnamefont {K.}~\bibnamefont
  {Lee}}, \bibinfo {author} {\bibfnamefont {A.}~\bibnamefont {Pathak}},
  \bibinfo {author} {\bibfnamefont {I.~W.}\ \bibnamefont {Stewart}},\ and\
  \bibinfo {author} {\bibfnamefont {Z.}~\bibnamefont {Sun}},\ }\href@noop {} {\
  }\Eprint {https://arxiv.org/abs/to appear} {to appear} \BibitemShut {NoStop}%
\bibitem [{\citenamefont {Chen}\ \emph
  {et~al.}(2023{\natexlab{b}})\citenamefont {Chen}, \citenamefont {Lee},
  \citenamefont {Chen}, \citenamefont {Chang}, \citenamefont {McGinn},
  \citenamefont {Sheng}, \citenamefont {Innocenti},\ and\ \citenamefont
  {Maggi}}]{Chen:2023nsi}%
  \BibitemOpen
  \bibfield  {author} {\bibinfo {author} {\bibfnamefont {Y.-C.}\ \bibnamefont
  {Chen}}, \bibinfo {author} {\bibfnamefont {Y.-J.}\ \bibnamefont {Lee}},
  \bibinfo {author} {\bibfnamefont {Y.}~\bibnamefont {Chen}}, \bibinfo {author}
  {\bibfnamefont {P.}~\bibnamefont {Chang}}, \bibinfo {author} {\bibfnamefont
  {C.}~\bibnamefont {McGinn}}, \bibinfo {author} {\bibfnamefont {T.-A.}\
  \bibnamefont {Sheng}}, \bibinfo {author} {\bibfnamefont {G.~M.}\ \bibnamefont
  {Innocenti}},\ and\ \bibinfo {author} {\bibfnamefont {M.}~\bibnamefont
  {Maggi}},\ }\bibfield  {title} {\bibinfo {title} {{Analysis note:
  two-particle correlation in $e^+e^-$ collisions at 91-209 GeV with archived
  ALEPH data}},\ }\href@noop {} {\  (\bibinfo {year} {2023}{\natexlab{b}})},\
  \Eprint {https://arxiv.org/abs/2309.09874} {arXiv:2309.09874 [hep-ex]}
  \BibitemShut {NoStop}%
\bibitem [{\citenamefont {Chang}\ \emph {et~al.}(2022)\citenamefont {Chang},
  \citenamefont {Kologlu}, \citenamefont {Kravchuk}, \citenamefont
  {Simmons-Duffin},\ and\ \citenamefont {Zhiboedov}}]{Chang:2020qpj}%
  \BibitemOpen
  \bibfield  {author} {\bibinfo {author} {\bibfnamefont {C.-H.}\ \bibnamefont
  {Chang}}, \bibinfo {author} {\bibfnamefont {M.}~\bibnamefont {Kologlu}},
  \bibinfo {author} {\bibfnamefont {P.}~\bibnamefont {Kravchuk}}, \bibinfo
  {author} {\bibfnamefont {D.}~\bibnamefont {Simmons-Duffin}},\ and\ \bibinfo
  {author} {\bibfnamefont {A.}~\bibnamefont {Zhiboedov}},\ }\bibfield  {title}
  {\bibinfo {title} {{Transverse spin in the light-ray OPE}},\ }\href
  {https://doi.org/10.1007/JHEP05(2022)059} {\bibfield  {journal} {\bibinfo
  {journal} {JHEP}\ }\textbf {\bibinfo {volume} {05}},\ \bibinfo {pages}
  {059}},\ \Eprint {https://arxiv.org/abs/2010.04726} {arXiv:2010.04726
  [hep-th]} \BibitemShut {NoStop}%
\bibitem [{\citenamefont {Caron-Huot}\ \emph {et~al.}(2023)\citenamefont
  {Caron-Huot}, \citenamefont {Kologlu}, \citenamefont {Kravchuk},
  \citenamefont {Meltzer},\ and\ \citenamefont
  {Simmons-Duffin}}]{Caron-Huot:2022eqs}%
  \BibitemOpen
  \bibfield  {author} {\bibinfo {author} {\bibfnamefont {S.}~\bibnamefont
  {Caron-Huot}}, \bibinfo {author} {\bibfnamefont {M.}~\bibnamefont {Kologlu}},
  \bibinfo {author} {\bibfnamefont {P.}~\bibnamefont {Kravchuk}}, \bibinfo
  {author} {\bibfnamefont {D.}~\bibnamefont {Meltzer}},\ and\ \bibinfo {author}
  {\bibfnamefont {D.}~\bibnamefont {Simmons-Duffin}},\ }\bibfield  {title}
  {\bibinfo {title} {{Detectors in weakly-coupled field theories}},\ }\href
  {https://doi.org/10.1007/JHEP04(2023)014} {\bibfield  {journal} {\bibinfo
  {journal} {JHEP}\ }\textbf {\bibinfo {volume} {04}},\ \bibinfo {pages}
  {014}},\ \Eprint {https://arxiv.org/abs/2209.00008} {arXiv:2209.00008
  [hep-th]} \BibitemShut {NoStop}%
\bibitem [{\citenamefont {Chen}\ \emph
  {et~al.}(2023{\natexlab{c}})\citenamefont {Chen}, \citenamefont {Gao},
  \citenamefont {Li}, \citenamefont {Xu}, \citenamefont {Zhang},\ and\
  \citenamefont {Zhu}}]{Chen:2023zlx}%
  \BibitemOpen
  \bibfield  {author} {\bibinfo {author} {\bibfnamefont {W.}~\bibnamefont
  {Chen}}, \bibinfo {author} {\bibfnamefont {J.}~\bibnamefont {Gao}}, \bibinfo
  {author} {\bibfnamefont {Y.}~\bibnamefont {Li}}, \bibinfo {author}
  {\bibfnamefont {Z.}~\bibnamefont {Xu}}, \bibinfo {author} {\bibfnamefont
  {X.}~\bibnamefont {Zhang}},\ and\ \bibinfo {author} {\bibfnamefont {H.~X.}\
  \bibnamefont {Zhu}},\ }\bibfield  {title} {\bibinfo {title} {{NNLL
  Resummation for Projected Three-Point Energy Correlator}},\ }\href@noop {} {\
   (\bibinfo {year} {2023}{\natexlab{c}})},\ \Eprint
  {https://arxiv.org/abs/2307.07510} {arXiv:2307.07510 [hep-ph]} \BibitemShut
  {NoStop}%
\bibitem [{\citenamefont {Chen}\ \emph
  {et~al.}(2024{\natexlab{a}})\citenamefont {Chen}, \citenamefont {Monni},\
  and\ \citenamefont {Zhu}}]{HaoTalkSCET}%
  \BibitemOpen
  \bibfield  {author} {\bibinfo {author} {\bibfnamefont {H.}~\bibnamefont
  {Chen}}, \bibinfo {author} {\bibfnamefont {P.}~\bibnamefont {Monni}},\ and\
  \bibinfo {author} {\bibfnamefont {H.~X.}\ \bibnamefont {Zhu}},\ }\bibfield
  {title} {\bibinfo {title} {Perturbative evolution of hadronization effects in
  energy correlators}} (\bibinfo {year} {2024}{\natexlab{a}}),\ \bibinfo {note}
  {talk by Hao Chen at SCET 2024}\BibitemShut {NoStop}%
\bibitem [{\citenamefont {Chen}\ \emph
  {et~al.}(2024{\natexlab{b}})\citenamefont {Chen}, \citenamefont {Monni},
  \citenamefont {Xu},\ and\ \citenamefont {Zhu}}]{Chen:2024nyc}%
  \BibitemOpen
  \bibfield  {author} {\bibinfo {author} {\bibfnamefont {H.}~\bibnamefont
  {Chen}}, \bibinfo {author} {\bibfnamefont {P.~F.}\ \bibnamefont {Monni}},
  \bibinfo {author} {\bibfnamefont {Z.}~\bibnamefont {Xu}},\ and\ \bibinfo
  {author} {\bibfnamefont {H.~X.}\ \bibnamefont {Zhu}},\ }\bibfield  {title}
  {\bibinfo {title} {{Scaling violation in power corrections to energy
  correlators from the light-ray OPE}},\ }\href@noop {} {\  (\bibinfo {year}
  {2024}{\natexlab{b}})},\ \Eprint {https://arxiv.org/abs/2406.06668}
  {arXiv:2406.06668 [hep-ph]} \BibitemShut {NoStop}%
\bibitem [{\citenamefont {Andersson}\ \emph {et~al.}(1983)\citenamefont
  {Andersson}, \citenamefont {Gustafson}, \citenamefont {Ingelman},\ and\
  \citenamefont {Sjostrand}}]{Andersson:1983ia}%
  \BibitemOpen
  \bibfield  {author} {\bibinfo {author} {\bibfnamefont {B.}~\bibnamefont
  {Andersson}}, \bibinfo {author} {\bibfnamefont {G.}~\bibnamefont
  {Gustafson}}, \bibinfo {author} {\bibfnamefont {G.}~\bibnamefont
  {Ingelman}},\ and\ \bibinfo {author} {\bibfnamefont {T.}~\bibnamefont
  {Sjostrand}},\ }\bibfield  {title} {\bibinfo {title} {{Parton Fragmentation
  and String Dynamics}},\ }\href {https://doi.org/10.1016/0370-1573(83)90080-7}
  {\bibfield  {journal} {\bibinfo  {journal} {Phys. Rept.}\ }\textbf {\bibinfo
  {volume} {97}},\ \bibinfo {pages} {31} (\bibinfo {year} {1983})}\BibitemShut
  {NoStop}%
\bibitem [{\citenamefont {Marchesini}\ \emph {et~al.}(1992)\citenamefont
  {Marchesini}, \citenamefont {Webber}, \citenamefont {Abbiendi}, \citenamefont
  {Knowles}, \citenamefont {Seymour},\ and\ \citenamefont
  {Stanco}}]{Marchesini:1991ch}%
  \BibitemOpen
  \bibfield  {author} {\bibinfo {author} {\bibfnamefont {G.}~\bibnamefont
  {Marchesini}}, \bibinfo {author} {\bibfnamefont {B.~R.}\ \bibnamefont
  {Webber}}, \bibinfo {author} {\bibfnamefont {G.}~\bibnamefont {Abbiendi}},
  \bibinfo {author} {\bibfnamefont {I.~G.}\ \bibnamefont {Knowles}}, \bibinfo
  {author} {\bibfnamefont {M.~H.}\ \bibnamefont {Seymour}},\ and\ \bibinfo
  {author} {\bibfnamefont {L.}~\bibnamefont {Stanco}},\ }\bibfield  {title}
  {\bibinfo {title} {{HERWIG: A Monte Carlo event generator for simulating
  hadron emission reactions with interfering gluons. Version 5.1 - April
  1991}},\ }\href {https://doi.org/10.1016/0010-4655(92)90055-4} {\bibfield
  {journal} {\bibinfo  {journal} {Comput. Phys. Commun.}\ }\textbf {\bibinfo
  {volume} {67}},\ \bibinfo {pages} {465} (\bibinfo {year} {1992})}\BibitemShut
  {NoStop}%
\bibitem [{\citenamefont {Bahr}\ \emph
  {et~al.}(2008{\natexlab{a}})\citenamefont {Bahr} \emph
  {et~al.}}]{Bahr:2008pv}%
  \BibitemOpen
  \bibfield  {author} {\bibinfo {author} {\bibfnamefont {M.}~\bibnamefont
  {Bahr}} \emph {et~al.},\ }\bibfield  {title} {\bibinfo {title} {{Herwig++
  Physics and Manual}},\ }\href
  {https://doi.org/10.1140/epjc/s10052-008-0798-9} {\bibfield  {journal}
  {\bibinfo  {journal} {Eur. Phys. J. C}\ }\textbf {\bibinfo {volume} {58}},\
  \bibinfo {pages} {639} (\bibinfo {year} {2008}{\natexlab{a}})},\ \Eprint
  {https://arxiv.org/abs/0803.0883} {arXiv:0803.0883 [hep-ph]} \BibitemShut
  {NoStop}%
\bibitem [{\citenamefont {Bahr}\ \emph
  {et~al.}(2008{\natexlab{b}})\citenamefont {Bahr}, \citenamefont {Gieseke},
  \citenamefont {Gigg}, \citenamefont {Grellscheid}, \citenamefont {Hamilton},
  \citenamefont {Platzer}, \citenamefont {Richardson}, \citenamefont
  {Seymour},\ and\ \citenamefont {Tully}}]{Bahr:2008tf}%
  \BibitemOpen
  \bibfield  {author} {\bibinfo {author} {\bibfnamefont {M.}~\bibnamefont
  {Bahr}}, \bibinfo {author} {\bibfnamefont {S.}~\bibnamefont {Gieseke}},
  \bibinfo {author} {\bibfnamefont {M.}~\bibnamefont {Gigg}}, \bibinfo {author}
  {\bibfnamefont {D.}~\bibnamefont {Grellscheid}}, \bibinfo {author}
  {\bibfnamefont {K.}~\bibnamefont {Hamilton}}, \bibinfo {author}
  {\bibfnamefont {S.}~\bibnamefont {Platzer}}, \bibinfo {author} {\bibfnamefont
  {P.}~\bibnamefont {Richardson}}, \bibinfo {author} {\bibfnamefont {M.~H.}\
  \bibnamefont {Seymour}},\ and\ \bibinfo {author} {\bibfnamefont
  {J.}~\bibnamefont {Tully}},\ }\bibfield  {title} {\bibinfo {title} {{Herwig++
  2.3 Release Note}},\ }\href@noop {} {\  (\bibinfo {year}
  {2008}{\natexlab{b}})},\ \Eprint {https://arxiv.org/abs/0812.0529}
  {arXiv:0812.0529 [hep-ph]} \BibitemShut {NoStop}%
\bibitem [{\citenamefont {Bellm}\ \emph {et~al.}(2020)\citenamefont {Bellm}
  \emph {et~al.}}]{Bellm:2019zci}%
  \BibitemOpen
  \bibfield  {author} {\bibinfo {author} {\bibfnamefont {J.}~\bibnamefont
  {Bellm}} \emph {et~al.},\ }\bibfield  {title} {\bibinfo {title} {{Herwig 7.2
  release note}},\ }\href {https://doi.org/10.1140/epjc/s10052-020-8011-x}
  {\bibfield  {journal} {\bibinfo  {journal} {Eur. Phys. J. C}\ }\textbf
  {\bibinfo {volume} {80}},\ \bibinfo {pages} {452} (\bibinfo {year} {2020})},\
  \Eprint {https://arxiv.org/abs/1912.06509} {arXiv:1912.06509 [hep-ph]}
  \BibitemShut {NoStop}%
\bibitem [{\citenamefont {Nachtmann}(1973)}]{Nachtmann:1973mr}%
  \BibitemOpen
  \bibfield  {author} {\bibinfo {author} {\bibfnamefont {O.}~\bibnamefont
  {Nachtmann}},\ }\bibfield  {title} {\bibinfo {title} {{Positivity constraints
  for anomalous dimensions}},\ }\href
  {https://doi.org/10.1016/0550-3213(73)90144-2} {\bibfield  {journal}
  {\bibinfo  {journal} {Nucl. Phys. B}\ }\textbf {\bibinfo {volume} {63}},\
  \bibinfo {pages} {237} (\bibinfo {year} {1973})}\BibitemShut {NoStop}%
\end{thebibliography}%

\end{document}